\RequirePackage{luatex85}
\PassOptionsToPackage{dvipsnames}{xcolor}
\RequirePackage{amssymb}
\RequirePackage{varioref}
\RequirePackage{fancyvrb}% must be before hyperref, see https://github.com/jgm/pandoc/issues/3361#issuecomment-274735020
\PassOptionsToPackage{nameinlink}{cleveref}%[2012/02/15]% v0.18.4;
\documentclass[sn-mathphys,pdflatex]{sn-jnl}% Math and Physical Sciences Reference Style
\catcode`|=12
\makeatletter
\usepackage{subcaption}
\usepackage[frozencache]{minted}
\usepackage{amsmath}
\usepackage{booktabs}
\usepackage{tabularx}
\usepackage{footnote}
\makesavenoteenv{tabular}
\makesavenoteenv{table}
\usepackage{amssymb}
\usepackage[defaultcolor=magenta,final]{changes}
\makeatletter
\renewcommand{\todo}[2][]{\tikzexternaldisable\@todo[#1]{#2}\tikzexternalenable}
\makeatother
\ifthenelse{\boolean{Changes@optiondraft}}{%
  \newcommand{\replacedboxed}[2]{\replaced{\mbox{#1}}{\mbox{#2}}}
}{%
  \let\replacedboxed=\replaced%}[2]{\replaced{#1}{#2}
}
\usepackage[all]{xy}
\usepackage{stmaryrd}
\usepackage[utf8]{inputenc}
\usepackage{textgreek}
\usepackage{newunicodechar}
\usepackage{usebib}
\usepackage{tikz}
\usepackage{pgfplots}
\pgfplotsset{compat=1.18}
\usepackage{multicol}
\usepackage{apptools}
\usepackage{placeins}
\usepackage{adjustbox}
\usepackage{xparse}
\usepackage{cleveref}
\usepgfplotslibrary{units}
\usepgfplotslibrary{external}
\tikzsetfigurename{rewriting-figure}
\newunicodechar{→}{\ensuremath{\to}}
\newunicodechar{⇒}{\ensuremath{\Rightarrow}}
\newunicodechar{λ}{\ensuremath{\lambda}}
\newunicodechar{∀}{\ensuremath{\forall}}
\newunicodechar{ℕ}{\ensuremath{\mathbb{N}}}

\newcommand{\autocommanameref}[1]{\autoref{#1}, \nameref{#1}}

\usepackage{etoolbox,xpatch}

\providebool{arxiv}
\booltrue{arxiv}

\@ifpackageloaded{breakurl}{
  \let\@badhref=\href
  \renewcommand{\href}[2]{\@badhref{#2}{#1}}
  \newcommand{\hreftt}[2]{\texttt{\href{#1}{#2}}}
}{
  \newcommand{\hreftt}[2]{\href{#1}{\texttt{#2}}}
}

\makeatletter
\AtBeginEnvironment{minted}{\dontdofcolorbox}
\def\dontdofcolorbox{\renewcommand\fcolorbox[4][]{##4}}
\xpatchcmd{\inputminted}{\minted@fvset}{\minted@fvset\dontdofcolorbox}{}{}
\xpatchcmd{\mintinline}{\minted@fvset}{\minted@fvset\dontdofcolorbox}{}{} % see https://tex.stackexchange.com/a/401250/
\newcommand{\textmintinline}[2]{\text{\mintinline{#1}{#2}}}
\makeatother

\makeatletter
\patchcmd{\hyper@makecurrent}{%
    \ifx\Hy@param\Hy@chapterstring
        \let\Hy@param\Hy@chapapp
    \fi
}{%
    \iftoggle{inappendix}{%true-branch
        \@checkappendixparam{chapter}%
        \@checkappendixparam{section}%
        \@checkappendixparam{subsection}%
        \@checkappendixparam{subsubsection}%
        \@checkappendixparam{paragraph}%
        \@checkappendixparam{subparagraph}%
    }{}%
}{}{\show\hyper@makecurrent\errmessage{failed to patch}}

\newcommand*{\@checkappendixparam}[1]{%
    \def\@checkappendixparamtmp{#1}%
    \ifx\Hy@param\@checkappendixparamtmp
        \let\Hy@param\Hy@appendixstring
    \fi
}
\makeatletter

\newtoggle{inappendix}
\togglefalse{inappendix}

\apptocmd{\appendix}{\toggletrue{inappendix}}{}{\show\appendix\errmessage{failed to patch}}
\pgfplotsset{
    discard if not/.style 2 args={
        x filter/.code={
            \edef\tempa{\thisrow{#1}}
            \edef\tempb{#2}
            \ifx\tempa\tempb
            \else
                
            \fi
        }
    }
}

\newbibfield{author}
\newbibfield{title}
\newbibfield{chapter}
\newbibfield{citet}
\bibinput{rewriting-lower}
\newcommand{\citetfield}[1]{\usebibentry{#1}{citet}~\cite{#1}}
\newcommand{\citetitle}[1]{\emph{\usebibentry{#1}{title}}~\cite{#1}}
\newcommand{\citechapter}[1]{\emph{\usebibentry{#1}{chapter}}~\cite{#1}}
\newcommand{\coqbug}[2][Coq bug~]{\href{https://github.com/coq/coq/issues/#2}{#1\##2}}
\newcommand{\coqpr}[2][Coq PR~]{\href{https://github.com/coq/coq/pull/#2}{#1\##2}}

\NewDocumentCommand{\githublink}{m m o}{%
  \IfNoValueTF{#3}%
    {\hreftt{https://github.com/#1/#2}{#1/#2}}%
    {\hreftt{https://github.com/#1/#2/tree/#3}{#1/#2@#3}}%
}
\NewDocumentCommand{\githublinkurl}{m m o}{%
  \IfNoValueTF{#3}%
    {\url{https://github.com/#1/#2}}%
    {\url{https://github.com/#1/#2/tree/#3}}%
}
\newcommand{\Rtac}{\ensuremath{\mathcal{R}_\text{\emph{tac}}}}
\newcommand{\Ltac}{\ensuremath{\mathcal{L}_\text{\emph{tac}}}}
\newcommand{\tacvmcompute}{\mintinline{coq}{vm_compute}}

\newcommand{\taccbv}{\mintinline{coq}{cbv}}

\newcommand{\tacsetoidrewrite}{\mintinline{coq}{setoid_rewrite}}
\newcommand{\tacrewritestrat}{\textbf{\texttt{rewrite\_strat}}}
\newcommand{\defeq}{=}
\makeatletter
\newcommand{\letin}[1][{\ensuremath{\cdots}}{\ensuremath{\cdots}}]{%
  \texttt{let }\@firstoftwo#1\texttt{ in }\@secondoftwo#1
}

\newcommand{\beginTikzpictureStamped}[2][]{%
  {%
    \everyeof{\noexpand}% IDK why \noexpand is the magic one, but I got it from http://mirrors.ibiblio.org/CTAN/macros/latex/contrib/oberdiek/catchfile.pdf
    \long\xdef\@tikzstamp{#2}%
  }%
  \def\@dobegintikzpicture{\begin{tikzpicture}[#1]}%
  \expandafter\@dobegintikzpicture\expandafter\def\expandafter\tikzstamp\expandafter{\@tikzstamp}%
}
\newcommand{\TikzpictureStamped}[3][]{\beginTikzpictureStamped[#1]{#2}#3\end{tikzpicture}}%
\makeatother
\newcommand{\asserteq}[2]{\ifthenelse{\equal{#1}{#2}}{}{\GenericError{}{Not equal: \detokenize{#1} != \detokenize{#2}}{}{}}}

\input{rewriting.arxiv-aux}
\newcommand{\arxivautonameref}[1]{\autoref*{arxiv:#1}, ``\nameref*{arxiv:#1}''}
\newcommand{\arxivautoreflong}[1]{\arxivautonameref{#1}, in the arXiv version of our ITP submission~\cite{RewritingITP2022}%
\let\arxivautoref\arxivautorefshort}
\newcommand{\arxivautorefshort}[1]{\arxivautonameref{#1}, in~\cite{RewritingITP2022}}
\let\arxivautoref\arxivautoreflong
\jyear{2023}%

\theoremstyle{thmstyleone}%
\theoremstyle{thmstyletwo}%

\theoremstyle{thmstylethree}%

\begin{document}
\VerbatimFootnotes

%\title[Article Title]{Article Title}
\title[Towards a Scalable Proof Engine]{Towards a Scalable Proof Engine: \mbox{\large A~Performant Prototype Rewriting Primitive for Coq}}

%%=============================================================%%
%% Prefix	-> \pfx{Dr}
%% GivenName	-> \fnm{Joergen W.}
%% Particle	-> \spfx{van der} -> surname prefix
%% FamilyName	-> \sur{Ploeg}
%% Suffix	-> \sfx{IV}
%% NatureName	-> \tanm{Poet Laureate} -> Title after name
%% Degrees	-> \dgr{MSc, PhD}
%% \author*[1,2]{\pfx{Dr} \fnm{Joergen W.} \spfx{van der} \sur{Ploeg} \sfx{IV} \tanm{Poet Laureate}
%%                 \dgr{MSc, PhD}}\email{iauthor@gmail.com}
%%=============================================================%%

%\todo{funding is supposed to be listed here, according to \url{https://www.springer.com/journal/10817/submission-guidelines\#Instructions\%20for\%20Authors\_Text}: ``Acknowledgments of people, grants, funds, etc. should be placed in a separate section on the title page. The names of funding organizations should be written in full.''}
\newcommand{\versionofrecordtext}{%
This version of the article has been accepted for publication, after peer review, but is not the Version of Record and does not reflect post-acceptance improvements / corrections.
Version of Record: \url{https://dx.doi.org/10.1007/s10817-024-09705-6}.
}
\newcommand{\fundingtext}{%
This work was supported in part by a Google Research Award, National Science Foundation grants CCF-1253229, CCF-1512611, and CCF-1521584, and the National Science Foundation Graduate Research Fellowship under Grant Nos.\ 1122374 and 1745302.
Any opinion, findings, and conclusions or recommendations expressed in this material are those of the authors and do not necessarily reflect the views of the National Science Foundation.
%GPT-4 was used to assist in the wordsmithing of this document.
% CCF-1253229 is CAREER: A Formal Verification Platform Focused on Programmer Productivity, Bedrock, 2013-06-01 through 2018-05-31
% CCF-1512611 is SHF: Medium: Fiat: Correct-by-Construction and Mostly Automated Derivation of Programs with an Interactive Theorem Prover https://www.nsf.gov/awardsearch/showAward?AWD_ID=1512611, September 1, 2015 to August 31, 2020 (Estimated),
% CCF-1521584 is Collaborative Research: Expeditions in Computing: The Science of Deep Specification, December 15, 2015 through November 30, 2020
% 1122374 and 1745302 are Jason's NSF GRFP Grant
% Google Research Award first shows up in commit 6cba11c7b09e7c3afbd7db2956adb47c78878738 by Adam in the reification by parametricity paper, and has been copied over from there
\ifbool{arxiv}{\versionofrecordtext}{}%
}

\expandafter\def\csname X@3\endcsname{3}%
\expandafter\def\csname X@4\endcsname{4}%

\author*[1,2]{\fnm{Jason} \sur{Gross}}\email{jgross@mit.edu}
\author[1,3]{\fnm{Andres} \sur{Erbsen}}\email{andreser@mit.edu}
\author[1,3]{\fnm{Jade} \sur{Philipoom}}\email{jade.philipoom@gmail.com}
\author[4]{\fnm{Rajashree} \sur{Agrawal}}\email{rajashree.agrawal@gmail.com}
\author[1]{\fnm{Adam} \sur{Chlipala}}\email{adamc@csail.mit.edu}

\affil[1]{\mbox{\orgdiv{CSAIL}, \orgname{Massachusetts Institute of Technology}, \orgaddress{%\street{32 Vassar St},
    \city{Cambridge}, \state{MA}% \postcode{02139}%
    , \country{USA}}}}

\affil[2]{\orgname{Machine~Intelligence~Research~Institute},~\orgaddress{\city{Berkeley},~\state{CA}%~\postcode{94704}%
    ,~\country{USA}}}

%\affil[3]{\orgname{Google}, \orgaddress{%\street{355 Main St},
%    \city{Cambridge}, \state{MA}% \postcode{02142}%
%    , \country{USA}}}
%\expandafter\show\csname X@4\endcsname
%\expandafter\show\csname X@6\endcsname
\footnotetext[3]{Now at Google}
\makeatletter
\let\save@thefootnote\thefootnote
\renewcommand{\thefootnote}{\fnsymbol{footnote}}
\footnotetext[2]{\fundingtext\vspace{-\baselineskip}}
\let\thefootnote\save@thefootnote
\makeatother
%\affil[4]{\orgname{Google Germany GmbH}, \orgaddress{%\street{ABC-Stra{\ss}e 19},
%    %\postcode{20354}
%    \city{Hamburg},
%    \country{Germany}}}

\affil[4]{\orgname{Reed~College},~\orgaddress{%\street{3203~SE~Woodstock~Blvd},~
      \city{Portland},~\state{OR}%~\postcode{97202}%
      ,~\country{USA}}}

%%==================================%%
%% sample for unstructured abstract %%
%%==================================%%

%\abstract{The abstract serves both as a general introduction to the topic and as a brief, non-technical summary of the main results and their implications. Authors are advised to check the author instructions for the journal they are submitting to for word limits and if structural elements like subheadings, citations, or equations are permitted.}
\abstract{
We address the challenges of scaling verification efforts to match the increasing complexity and size of systems.
We propose a research agenda aimed at building a performant proof engine by studying the asymptotic performance of proof engines and redesigning their building blocks.
As a case study, we explore equational rewriting and introduce a novel prototype proof engine building block for rewriting in Coq, utilizing proof by reflection for enhanced performance.

Our prototype implementation can significantly improve the development of verified compilers, as demonstrated in a case study with the Fiat Cryptography toolchain.
The resulting extracted command-line compiler is about 1000$\times$ faster while featuring simpler compiler-specific proofs.
This work lays some foundation for scaling verification efforts and contributes to the broader goal of developing a proof engine with good asymptotic performance, ultimately aimed at enabling the verification of larger and more complex systems.
%  Compilers are a prime target for formal verification, since compiler bugs invalidate higher-level correctness guarantees, but compiler changes may become more labor-intensive to implement, if they must come with proof patches.
%  One appealing approach is to present compilers as sets of algebraic rewrite rules, which a generic engine can apply efficiently.
%  Now each rewrite rule can be proved separately, with no need to revisit past proofs for other parts of the compiler.
%  We present the first realization of this idea, in the form of a framework for the Coq proof assistant.
%  Our new Coq command takes normal proved theorems and combines them automatically into fast compilers with proofs.
%  We applied our framework to improve the Fiat Cryptography toolchain for generating cryptographic arithmetic, producing an extracted command-line compiler that is about 1000$\times$ faster while actually featuring simpler compiler-specific proofs.
}


%%\pacs[JEL Classification]{D8, H51}

%%\pacs[MSC Classification]{35A01, 65L10, 65L12, 65L20, 65L70}
\ifbool{arxiv}{\pretocmd{\abstracthead}{\vspace*{-35pt}}{}{\show\abstracthead\errmessage{failed to patch}}}{}%

\maketitle

\section{Introduction}\label{sec:intro}

This paper is about verification at scale.
We investigated and worked on scalable verification concretely via improving Fiat Cryptography~\cite{FiatCryptoSP19} which generates code for big-integer modular arithmetic at the heart of elliptic-curve-cryptography algorithms.
Routines generated (with proof) with Fiat Cryptography now ship with all major Web browsers and all major mobile operating systems.
However, a paper about scaling would be lacking if limited to the scale of current projects instead of the scale of possible projects.

\subsection{Blue Sky}\label{sec:sky}

Let us ask an ambitious question:
\textbf{What would it take for verification to keep pace with the scale of the largest computations being run?}

Any less ambitious performance goal for automated proof tooling can be expended to fall short of real-world use cases where rigorous verification would be appreciated.
Across projects, verification typically involves a 10$\times$ -- 100$\times$ overhead in lines of verification over lines of code being verified~\cite{Gross2021thesis}.
The largest of current verification projects are a little more than $10^6$ lines of verification code.
This overhead, of course, is limiting the applicability of formal verification, so it makes sense to look at use cases that have not yet been tackled.

The standard list of software bug disasters cited in verification papers includes
Intel Pentium's \$475M chip recall over a floating point division bug~\cite{Infamous2014Gawron,Truth1995Halfhill,Pentium2011Nicely},
the crashing of NASA's \$235M Mars Climate Orbiter~\cite{lynch2017worst_mars,Metric1999Lloyd,Space1996Leech},
Therac-25 radiation overdoses~\citep[p.434]{Gift2018Baase}\cite{lynch2017worst_therac,Investigation1993Leveson},
the loss of the \$370M Ariane 5 rocket~\cite{lynch2017worst_ariane,Design1997Jazequel,Ariane1997Dowson},
and Knight Capital's loss of \$460M in 45 minutes due to a configuration bug~\cite{lynch2017worst_460m}.
Scaling verification to similar systems to prevent future mishaps will require qualitatively improving the peformance characteristics of verification tools.

But one may wish to aim even higher: Large Language Models (LLMs) are part of a young and vibrant field, and gaining correctness guarantees about them is both obviously desirable and an open, nebulous problem.
These ``programs'' consist of a small handful of relatively simple matrix operations arranged in standard ways, operating on truly massive matrices.
The size of these programs is measured by the number of parameters of the model; each entry in a matrix is a single parameter.
LLM size has grown at a rate of approximately $10\times$ \emph{per year}, increasing from around 94 million parameters in 2018 to over 500 billion in 2021~\cite{smith2022using}.
Wildly extrapolating from current projects, verifying an LLM would take $10^{12}$--$10^{13}$ lines of verification code.
This would require increasing our largest efforts thus far by at least 6 orders of magnitude!

Of course, there are a number of challenges in understanding, let alone verifying, machine-learning-derived artifacts -- but this does not appear to be an issue for practical deployment and subsequent consequences, so it would be dissatisfying to just give up!
So far, the reported consequences of occasional incorrect operation of these systems have been indirect and thus hard to compare, but they already cover a wide range from suicide~\cite{xiang2023-ai-suicide} to slander~\cite{turley2023chatgpt} to corporate embarrassment~\cite{lee2016learning, hern2016microsoft} and \$100B valuation impact~\cite{coulter2023alphabet}.
Even if we leave aside fuzzy specifications like ``never encourages harming anyone'' or ``provides only helpful and truthful responses, except when doing so would encourage harm'' and just seek to verify basic properties of LLMs that would be feasibly ascertainable about a conventional program, the sheer size of LLMs is a non-starter by itself.
For example, we can still seek to prove that ``basic arithmetic expressions of 20 characters or fewer are always accurately completed''\footnote{%
More formally, any query consisting of 19 or fewer characters from the string \verb|0123456789+-()*/^|, followed by \texttt{=}, is completed with its correct evaluation whenever the query is a valid arithmetic expression that computes to an integer.%
}
or, more ambitiously, ``every response to any query contains an `end-of-sequence' delimiter within the first 32k tokens'' or ``for any response $\mathcal R$ ever given by the model, the completion of the query `Answer with a single word, either Yes or No: Is the following text hate-speech? $\mathcal R$' will always be \mbox{`No.'}''
But for goals of this kind, trustworthy and efficient formal analysis of comparably large programs appears to be a strict prerequisite.

When looking to scale computer-checked proofs, there are two basic approaches available to us: we can produce a machine-checkable proof, or we can write a computational validator that we trust or prove correct.
The latter approach shines when the domain of reasoning is well-delimited: SMT solvers, for example, are great on domains that they naturally support.
But when the domain of reasoning is open and unbounded---as in any evolving, real-world product such as compilers, microkernels, file systems, hardware designs---we need machine-checkable proofs, likely relying on proven-correct or proof-producing domain-specific validators whenever appropriate.

While hand-written machine-checkable proofs may suffice for some of the applications currently targeted by verification, we cannot hope to scale such effort as we aim ourselves at larger and larger applications.
We need automation to generate the proofs for us.

The framework atop which we build automatic generation of machine-checkable proofs is a \emph{proof engine}.
A proof engine consists of \emph{proven correct} or trusted building blocks that can be \emph{modularly} combined to form proofs, and this set of building blocks must be \emph{powerful or complete} enough to prove any property we might want to validate.

Naively, automated approaches should scale linearly with input size.
The trouble is that proof engine \emph{performance} often scales exponentially in the size of programs being verified~\cite{Gross2021thesis}.
For example, as we'll see in \autoref{sec:fiat-crypto-slowness}, exponential scaling means that proof automation that succeeds in seconds on toy examples would not complete within the lifetime of the universe on real-world examples!
Then the grand target is
\textbf{A proof engine with good asymptotic performance.}

\subsection{Performant Rewriting}\label{sec:approach}

We see two sets of research priorities for building a performant proof engine:

\begin{itemize}
	\item Studying proof engines:
	What is the \emph{asymptotic performance} of proof engines?
	What's fast, what's slow, and why?
	What algorithms would get around the slowness?
	\item Redesigning proof engine building blocks:
	Once we have algorithms that are asymptotically faster than existing proof engines, what do we need to close the gap between ``having a performant algorithm'' and getting an actually usable proof engine building block?
\end{itemize}

We took a stab at this research for equational rewriting, an essential technique for simplifying expressions, proving equivalence between programs, and streamlining the development of verified systems.
Among the 70 or so developments tested in the Continuous Integration of Coq, our target proof assistant, the \mintinline{coq}{rewrite} tactic occurs over 350,000 times, being surpassed only by \mintinline{coq}{apply}.

Unfortunately, it appears that building blocks provided by current proof engines do not allow for a performant direct implementation of rewriting on their data structures.
In this paper, we study the underpinnings of the asymptotic performance inadequacies of equational rewriting built atop existing proof engines, and demonstrate how to use proof by reflection to integrate an efficient implementation of a rewriting proof engine building block.
%to an existing proof engine with support for fast computation.
%so we used proof by reflection to integrate an implementation that uses data structres we designed for this purpose
%we studied the asymptotic performance inadequacies of equational rewriting built atop existing proof engines, and ultimately designed
%and redesigning the proof engine building block for rewriting to accommodate performance strategies utilizing proof by reflection.

Proof by reflection is a well-known strategy for performance, enabling leveraging the efficiency of writing dedicated computational verifiers within general purpose proof assistants.
Once we write and prove correct an efficient computation that checks a given property holds, we can combine the correctness proof with the primitives for fast full reduction shipped in Coq's kernel, via a virtual machine~\cite{vmcompute} or compilation to native code~\cite{nativecompute}.

We presented our prototype implementation of a \mintinline{coq}{rewrite} building block in \citetitle{RewritingITP2022}, where we described our design, implementation, and performance evaluation.
In that paper we introduced our framework as a novel and powerful tool for formally verified rewriting in Coq, demonstrating its potential to improve the development of verified compilers, like CompCert~\cite{Compcert} and CakeML~\cite{CakeML}, through a case study with Fiat Cryptography.
In this extended version, we explore our prototype implementation in context of building performant proof engines.
In \replacedboxed{\cref{sec:status-quo-inadequate,sec:why-status-quo-bad}}{\autoref{sec:status-quo-inadequate}, \autoref{sec:why-status-quo-bad}} we investigate the inadequacy of the status quo for rewriting and provide explanations for contributing factors.
In \autoref{sec:better-than-status-quo} we lay out a desiderata for adequate proof engine building block for rewriting, against which we evaluate our prototype in \autoref{sec:our-prototype}.
Then we reprise the content of \citetitle{RewritingITP2022} in \cref{sec:structure,sec:scaling,sec:evaluation}.
Finally, we lay out a research agenda in \autoref{sec:future-work} for developing our prototype into an adequately performant, modular, and complete proof engine building block.

The paper focuses primarily on dependently typed proof assistants like Coq.
Where our limited expertise allows, we make some comparisons with Isabelle/HOL.
A more thorough evaluation of performant proof engine building blocks in other proof assistants might be useful.

%\section{Prototyping an Adequate Proof Engine Building Block for Rewriting}\label{sec:prototype}

\section{Where is the Status Quo on Rewriting?}\label{sec:status-quo-inadequate}\label{sec:fiat-crypto-slowness}

Many proof assistants, Coq included, ship with equational rewriting tactics.
Let us start by evaluating the rewriting tactics on Fiat Cryptography, a real-world project that generates 10s -- 100s lines of code per function.

Fiat Cryptography uses rewriting and partial evaluation to specialize generic templates for performing big-integer modular arithmetic to specific primes and machine architectures.
The generic templates are written as functional programs which can be both implemented and proven correct with just a handful of lines of code and proof.
The code output is straightline C (or Go, Rust, Zig, etc.), avoiding loops, branching constructs, and memory allocation; these demands are used to meet the requirement that code used to implement internet security be efficient and timing-side-channel-free.

The relevant parameter for performance scaling evaluation in Fiat Cryptography is the number of \emph{limbs}: modular arithmetic of big integers is performed by carving each big integer up into multiple 32- or 64-bit \emph{machine words}.
The number of machine words used to represent a single big integer is the number of \emph{limbs}.
For any given arithmetic template, the total number of lines of code in the output of the Fiat Cryptography compiler %, and thus the total number of binders in the term representation,
is determined almost entirely by the number of limbs used.
The smallest of toy examples might use one or two limbs; widely-used Curve25519 implementations use 5 limbs on 64-bit machines (10 limbs on 32-bit machines) to represent numbers up to $2^{255}-19$; P-256 uses 4 limbs (or 8 limbs) for numbers up to around $2^{256}$; P-384 uses 6 limbs (or 12 limbs) for numbers up to around $2^{384}$.
Our largest example, P-521, uses 9 limbs (or 18 limbs) to represent numbers up to around $2^{521}$.

%In order to compare performance of our prototype rewriting engine against existing engines,
% to run partial evaluation on the Fiat Cryptography arithmetic templates.
When attempting to use \tacsetoidrewrite{} for partial evaluation and rewriting on unsaturated Solinas on a prime requiring 4 limbs, we ran into an out-of-memory error after using over 60 GB RAM!
See \coqbug{13576} for more details and for updates.
Then we painstakingly optimized typeclass instances and rewriting lemmas so that we could use \tacrewritestrat{} instead, which does not require duplicating the entire goal at each rewriting step.
We arrived at invocation involving \emph{sixteen} consecutive calls to \tacrewritestrat{} with varying arguments and strategies.
While we were able to get up to 4 limb examples, extrapolating from the exponential asymptotics of the fastest growing subcall to \tacrewritestrat{} indicates that our smallest real-world example of 5 limbs would take 11 hours, and our largest real-world example of 17 limbs would take over 1000$\times$ the age of the universe.
See \autoref{tab:setoid-rewrite-perf-nlimbs} and see \coqbug{13708} for more details and updates.

\begin{figure}
\begin{tabular}{ | c | c | c | c| }
  \hline
  \# limbs & \tacsetoidrewrite{} &  \tacrewritestrat{} & extrapolated \tacrewritestrat{}  \\
  \hline
    1& 100 s & 11 s & \\
    \hline
    2& 10 m & 90 s  & \\
     \hline
    3& 3.5 h & 10 m & \\
     \hline
    4& out of memory & 70 m & \\ %17 GB RAM for 70 mins rewrite_strat
      \hline
    5&  & & 11 h \\
     \hline
    6&  & & 10 days \\
     \hline
    7&  & & 32 weeks \\
     \hline
    8&  & & 13 years \\
     \hline
    9&  & & 2 centuries \\
     \hline
    10&  & & 6 millennia \\
    \hline
    15&  & & 3$\times$ the age of the universe \\
      \hline
    17 &  & & 1000$\times$ the age of the universe \\
  \hline
  \end{tabular}
\caption{\label{tab:setoid-rewrite-perf-nlimbs}Performance numbers for \tacsetoidrewrite{} and \tacrewritestrat{}}
\end{figure}

We also tried rewriting in Lean in the hopes that a proof \replaced{assistant}{asisstant} specifically optimized for performance would be up to the challenge.
Although Lean performed about 30\% better than Coq's \tacsetoidrewrite{} on the 1-limb example, taking a bit under a minute, it did not complete on the two-limb example even after four hours (after which we stopped trying), and a five-limb example was still going after 40 hours.

In \autoref{fig:fiat-crypto-scaling-with-setoid} we will see that using our rewriting prototype for Fiat Cryptography results in a tool that can synthesize code for 32-bit P-521 in under four minutes.

\section{Why is the Status Quo so Slow?}\label{sec:why-status-quo-bad}\label{sec:setoid-rewrite-bottlenecks}

%The Practical Performance Bottlenecks of Proof-Producing Rewriting

%Although we have made our performance comparison against the built-in Coq tactics \tacsetoidrewrite{} and \tacrewritestrat{}, by analyzing the performance in detail, we can argue that these performance bottlenecks are likely to hold for any proof assistant designed like Coq.

In Fiat Cryptography, we ease proving correctness of the arithmetic templates by using a shallowly embedded representation, implementing cryptographic primitives as functions in Coq's functional programming language Gallina.
This shallow embedding forces us to encode subterm sharing using \letin{} binders, one binder for each variable assignment.
All of the rewriting performance bottlenecks we encountered that scale superlinearly in the number lines of code result from underlying superlinear scaling of rewriting in the number of binders.

Detailed debugging reveals six performance bottlenecks in the existing rewriting tactics (\mintinline{coq}{rewrite}, \tacsetoidrewrite{}, \tacrewritestrat{}) in Coq.
These bottlenecks contribute to the scaling we see in the microbenchmark of \autoref{fig:timing-UnderLetsPlus0}, which is explained in more detail in \autoref{sec:micro} and in complete detail in \arxivautoreflong{sec:UnderLetsPlus0-more}.

\subsection{Inefficient Matching Representations involving Existential-Variable Contexts}

We found that even when there are no occurrences fully matching a given rewrite rule, \tacsetoidrewrite{} can still be \emph{cubic} in the number of binders (or, more accurately, quadratic in the number of binders with an additional multiplicative linear factor of the number of head-symbol matches).
It is easy to end up in a situation where the majority of time in \tacsetoidrewrite{} is spent dealing with partial matches that ultimately fail to match.
%Rewriting without any successful matches takes nearly as much time as \tacsetoidrewrite{} in this microbenchmark.
%By the time we are looking at goals with 400 binders, the difference is less than 5\%.

We posit that the overhead in this microbenchmark comes from \tacsetoidrewrite{} looking for head-symbol matches and then creating evars (existential variables) to instantiate the arguments of the lemmas for each head-symbol-match location.
Even if there are no matches of the rule as a whole, there may still be head-symbol matches!

Coq uses a locally nameless representation~\cite{LocallyNameless} for its terms, so evar contexts are necessarily represented as \emph{named} contexts.
Representing a substitution between named contexts takes linear space, even when the substitution is trivial, resulting in each evar incurring linear overhead in the number of binders above it.
Furthermore, fresh-name generation in Coq is quadratic in the size of the context, and since evar-context creation uses fresh-name generation, the additional multiplicative factor likely comes from fresh-name generation~\cite{coqbug-12524}.

To eliminate the overhead in the microbenchmark, Coq would likely have to represent identity evar contexts using a compact representation, which is only naturally available for de Bruijn representations.%
\footnote{%
See \coqbug{12526} for updates.%
}
Any rewriting system that uses unification variables with a locally nameless (or named) context will incur at least quadratic overhead on this benchmark!

Note that \tacrewritestrat{} uses exactly the same rewriting engine as \tacsetoidrewrite{}, just with a different strategy.
We found that \tacsetoidrewrite{} and \tacrewritestrat{} have identical performance when there are no matches and generate identical proof terms when there are matches.
Hence we conclude that the difference in performance between \tacrewritestrat{} and \tacsetoidrewrite{} is entirely due to an increased number of failed rewrite attempts.

\subsection{Proof-Term Size}

Setting aside the performance bottleneck in constructing the matches in the first place, we can ask the question: how much cost is associated to the proof terms?
One way to ask this question in Coq is to see how long it takes to run \mintinline{coq}{Qed}.
While \mintinline{coq}{Qed} time is asymptotically better than proof construction time, it is still quadratic in the number of binders.
This outcome is unsurprising, because the proof-term size is quadratic in the number of binders.
On this microbenchmark, we found that \mintinline{coq}{Qed} time hits one second at about 250 binders, and using the best-fit quadratic line suggests that it would hit 10 seconds at about 800 binders and 100 seconds at about 2\,500 binders.
While this may be reasonable for our microbenchmarks, which only contain as many rewrite occurrences as there are binders, it would become unwieldy to try to build and typecheck such a proof with a rule for every primitive reduction step, which would be required if we want to avoid manually converting the code in Fiat Cryptography to continuation-passing style.
%\todo{is this too much detail?}

The quadratic factor in the proof term comes because we repeat subterms of the goal linearly in the number of rewrites.
For example, if we want to rewrite \mintinline{coq}{f (f x)} into \mintinline{coq}{g (g x)} by the equation \mintinline{coq}{∀ x, f x = g x}, then we will first rewrite \mintinline{coq}{f x} into \mintinline{coq}{g x}, and then rewrite \mintinline{coq}{f (g x)} into \mintinline{coq}{g (g x)}.
Note that \mintinline{coq}{g x} occurs three times (and will continue to occur in every subsequent step).
Also note that the duplication appears already in the statements of the intermediate theorems being proven, so it would be a consideration regardless of the representation of the proofs themselves.

Although the \tacrewritestrat{} tactic makes some effort to avoid duplication in the proof term, doing much better than any of the other tactics shipped with Coq, it still produces proof terms that are far from optimal.
While multiple rewrites can be chained on subterms without having to duplicate the surrounding context, no effort is made to avoid duplicating subterms at different depths in the AST.

\subsection{Poor Subterm Sharing}

How easy is it to share subterms and create a linearly sized proof?
While it is relatively straightforward to share subterms using \mintinline{coq}{let} binders when the rewrite locations are not under any binders, it is not at all obvious how to share subterms when the terms occur under different binders.
Hence any rewriting algorithm that does not find a way to share subterms across different contexts will incur a quadratic factor in proof-building and proof-checking time, and we expect this factor will be significant enough to make applications to projects as large as Fiat Crypto infeasible.

\subsection{Asymptotic Performance Challenges in Rewriting}\label{sec:proof-producing-rewriting:theoretical-performance}

Even setting aside the sharing and typechecking problems of proof-term checking, there is still an asymptotic bottleneck in generating such a proof term, or implementing the rewriting in an incremental manner at all.
We specifically consider here a rewriting tactic which is built from smaller \emph{correct} and \emph{modular} primitives.
A good proof engine allows combining its primitive building blocks efficiently while still getting \emph{local} error messages about mistakes;
proving would be quite tricky if there was no feedback on incorrect proofs until \mintinline{coq}{Qed}-time!

Rather than considering a concrete microbenchmark in this sub\deleted{sub}section, we analyze asymptotic performance of the pseudocode of \Cref{fig:rw-app,fig:rw-lam}.

Consider the dataflow-directed rewriter \mintinline{coq}{rw}, parametrized by a local rewriting function \mintinline{coq}{rwh} (``rewrite head'').
Both \mintinline{coq}{rw} and \mintinline{coq}{rwh} and take as input a expression \mintinline{coq}{e} in which rewriting will be performed, and return either ``nothing to rewrite'' or an expression \mintinline{coq}{e'} and the theorem \mintinline{coq}{e' = e}.
The analysis here will only assume that the representation of a theorem allows its statement to be read out, allowing for instantions with explicit proof terms or LCF-style abstract types.
A simple rewrite-head function would match its input against the left-hand side of a quantified equality theorem, and return (an instantiated version of) the right-hand side and the theorem in case of success.
\mintinline{coq}{rw} simply \replaced{recurses}{recursies} over the structure of the expression, applying \mintinline{coq}{rwh} at each node after rewriting it recursively.
The application case appears in \Vref{fig:rw-app}.

\begin{figure}
%  \begin{subfigure}{\textwidth}
\begin{minted}[fontsize=\small]{coq}
  rw (f x) =
  let (mid, fx_mid) :=
     match rw f, rw x with
     | (f', f'f), (x', x'x) => (f' x', app_cong f'f x'x)
     | _ => (f x, eq_refl (f x))
     end in
   match rwh mid with
   | (result, mid_result) => (result, eq_trans fx_mid mid_result)
   | _ => (mid, fx_mid)
   end
\end{minted}
\caption{\label{fig:rw-app}Partial Pseudocode Rewriting: Function Application Case}
%  \end{subfigure}
\end{figure}

The intent of the code is clear as far as to which expressions and theorems it manipulates (returning \mintinline{coq}{r} and \mintinline{coq}{f x = r}), but to understand the performance we also need to pay attention to how these objects are represented.
As each theorem must at least include its statement, we have \emph{eight} occurrences of the function argument if both matches succeed: in the input expression (\mintinline{coq}{x}), the expression after recursive rewriting (\mintinline{coq}{x'}), the theorem that relates the two (\mintinline{coq}{x'x}), the \mintinline{coq}{f} applied to the same expressions and its equality proof (\mintinline{coq}{mid} and \mintinline{coq}{fx_mid}), the expression after rewriting in mid the head position (\mintinline{coq}{result}) and its equality proof (\mintinline{coq}{mid_result}), and the combined equality proof for the two rewriting steps (rw (f x)).
In this example, straightforward sharing of syntactic subexpressions using aliasing pointers can completely avoid storing multiple copies of the expression, relying on run-time automatic memory management as one would expect given the ML-like syntax in the example.
Both Coq and Lean implement this optimization.

However, maintaining sharing in memory is only the beginning: it is also important to avoid duplicate computation on the implicitly shared subexpressions.
The strategy in Lean is centered around memoizing functions commonly applied to terms where loss of sharing would result in dramatic slowdown, whereas Coq developers seek to minimize the use of functions that would be slow on expressions whose form without sharing is large.
In both cases, developments using the proof engine have come to rely on the speedups.
%For example, Coq 8.\todo{TODO} and later include a heuristic that skips checking of proofs constructed in a manner similar to the last example at some proof steps.
%This is sound if the proof was constructed in a context subsumed by the context in which it is used, but there is no check for that either.
%This leads to assertion failures (sometimes deep inside proof automation), user confusion, and the proof engine accepting bad proofs that are only rejected at \mintinline{coq}{Qed} when the accumulated proof term is checked again.
%Regardless, the heuristic has shipped in \todo{TODO} subsequent releases because the performance benefit is too large to give up.

We will return to deduplication of computation shortly, but first, let's analyze how the general strategy so far fares in expressions that contain binders in the expression itself.
The $\lambda$ abstraction case is shown in \Vref{fig:rw-lam}.

\begin{figure}
%  \begin{subfigure}{\textwidth}
\begin{minted}[fontsize=\small]{coq}
rw (λ x:T, e) =
   let rrw := (λ x:T, rw e) in
   let mid := (λ y:T, let (e', _) := beta (rrw y) in e') in
   let f_mid:=λ_extensionality(λ z:T, let (_,e'e):=beta (rrw z) in e'e) in
   match rwh mid with
   | (result, mid_result) => (result, eq_trans f_mid mid_result)
   | _ => (mid, f_mid)
   end
\end{minted}
\caption{\label{fig:rw-lam}Partial Pseudocode Rewriting: $\lambda$ Abstraction Case}
%  \end{subfigure}
%  \caption{\label{fig:rw}Partial Pseudocode Implementation of Rewriting}
\end{figure}

The challenge in this case is the representation of the two components of the return value: the expression and the theorem.
Unlike the function application case, \added{the} theorem is no longer constructed directly from the new expression, rather both are the result of beta-reducing different applications of the lambda under which the recursive rewriting is performed.
Depending on the underlying term representation, the two copies of the resulting expression may even not be represented using identical meta-language objects.
Further, the notation these examples elides ``lifting'' -- extracting a well-typed term in an initial context and then using it in a strictly extended context -- which is non-trivial in representations involving de Brujin indices.
Implementing this pseudocode using the Coq ML API would likely result the expression being copied 5 times: lifting \mintinline{coq}{e} over \mintinline{coq}{x}, lifting \mintinline{coq}{rrw} over \mintinline{coq}{y}, substituting \mintinline{coq}{y} into \mintinline{coq}{rrw}, lifting \mintinline{coq}{rrw} over \mintinline{coq}{z}, and substituting \mintinline{coq}{z} into \mintinline{coq}{rrw}.
It is possible to generate the same proof term in linear time, but we are not aware of any existing term representation where this can be achieved using a small set of generic primitive operations supporting dependent types whose soundness can be checked incrementally.

Note that if we drop the requirement for dependent type support, we \emph{could} implement such an algorithm in Isabelle/HOL.
Although rewriting suffers from a similar quadratic asymptotic blowup in the standard kernel, this blowup can be avoided when using a kernel based on nominal binders.
In Isabelle's nominal kernel, closing an open term over a binder is $\mathcal{O}(1)$:
an open nominal term can be closed without walking the term, and if variable references are scoped by type, no additional typechecking is needed either.
%because variable namespaces are segregated by type:
%\mintinline{coq}{x : int} and \mintinline{coq}{x : string} live in different namespaces, so constructing and checking \mintinline{coq}{λ x : int, body} from \mintinline{coq}{body} does not require walking \mintinline{coq}{body}.
This approach does not scale to dependently typed proof assistants, where types can include variables and where checking equality of types can be arbitrarily expensive.
In theory, the quadratic blowup arising in the standard kernel could be avoided if there were a primitive for simultaneously closing an open term over a collection of binders, walking the term only once.
Even this solution is not entirely satisfying, as the delaying of term closure ought to be an optimization performed by the tactic language, not by the implementer of \mintinline{coq}{rewrite} and similar tactics.
Furthermore, without appropriate support for tracking the provenance of open terms, delaying the closing would result in non-local errors.

\subsection{Empirical Cost of Incrementality}

In line with out analysis, the performance results reported by \citetfield{Aehlig} suggest that even if all of the superlinear bottlenecks were fixed---no small undertaking---rewriting and partial evaluation via reflection might still be orders of magnitude faster than any tactic that constructs theorems for intermediate results.
\citetfield{Aehlig} reported a $10\times$--$100\times$ speed-up of their rewriting tactic over the \emph{simp} tactic in Isabelle, which performs all of the intermediate rewriting steps via the kernel API.
Their rewriting tactic follows an approach relatively similar to ours (see \autoref{sec:aehlig-intro}), although they avoid producing proofs by stepping outside of Isabelle/HOL's TCB.
%For instance, while the reported performance experiments of \citetfield{Aehlig} generate only closed terms with no binders, Fiat Cryptography may generate a single routine (e.g., multiplication for curve P-384) with nearly a thousand nested binders.

\subsection{Overhead from the \texorpdfstring{\mintinline{coq}{let}}{let} Typing Rule}

Returning to concrete issues in Coq, suppose we had a proof-producing rewriting algorithm that shared subterms even under binders.
Would it be enough?
It turns out that even when the proof size is linear in the number of binders, the cost to typecheck it in Coq is still quadratic!
The reason is that when checking that \texttt{f : T} in a context \mintinline{coq}{x := v}, to check that \mintinline{coq}{let x := v in f} has type \texttt{T} (assuming that \mintinline{coq}{x} does not occur in \texttt{T}), Coq will substitute \mintinline{coq}{v} for \mintinline{coq}{x} in \texttt{T}.
So if a proof term has $n$ \mintinline{coq}{let} binders (e.g., used for sharing subterms), Coq will perform $n$ substitutions on the type of the proof term, even if none of the \mintinline{coq}{let} binders are used.
If the number of \mintinline{coq}{let} binders is linear in the size of the type, there is quadratic overhead in proof-checking time, even when the proof-term size is linear.

We performed a microbenchmark on a rewriting goal with no binders (because there is an obvious algorithm for sharing subterms in that case) and found that the proof-checking time reached about one second at about 2\,000 binders and reached 10 seconds at about 7\,000 binders.
While these results might seem good enough for Fiat Cryptography, we expect that there are hundreds of thousands of primitive reduction/rewriting steps even when there are only a few hundred binders in the output term, and we would need \mintinline{coq}{let} binders for each of them.
Furthermore, we expect that getting such an algorithm correct would be quite tricky.

Fixing this quadratic bottleneck would, as far as we can tell, require deep changes in how Coq is implemented; it would either require reworking all of Coq to operate on some efficient representation of delayed substitutions paired with unsubstituted terms, or else it would require changing the typing rules of the type theory itself to remove this substitution from the typing rule for \mintinline{coq}{let}.
Note that there is a similar issue that crops up for function application and abstraction.

\section{Desiderata for a New Rewriting Building Block}\label{sec:better-than-status-quo}

Existing built-in rewriting tactics fall short in addressing real-world project requirements.
Naturally, the field must look towards building a new rewriting building block that would be adequate for real-world projects.

There are two strategies for ensuring correctness of such a rewriting building block.
  \emph{Proof-producing} \mintinline{coq}{rewrite} tactics build proofs out of smaller, trusted or proven-correct building blocks.
  \emph{Proven correct} \mintinline{coq}{rewrite} tactics implement a procedure for computing the final term, and have an associated proof that whatever rewritten term they output will be equal to the initial term.

  \citetfield{Hickey2006} discuss at length how to build compilers around proof-producing rewrite rules.
``All program transformations, from parsing to code generation, are cleanly isolated and specified as term rewrites.''
While they note that the correctness of the compiler is thus reduced to the correctness of the rewrite rules, they did not prove correctness mechanically.
More importantly, it is not clear that they manage to avoid the asymptotic blow-up associated with proof-producing rewriting of deeply nested let-binders.
Since they give no performance numbers, it is hard to say whether or not their compiler performs at the scale necessary for Fiat Cryptography.
%Their rewrite-engine driver is unproven OCaml code, while we will produce custom drivers with Coq proofs.

  %
  %Although most existing \mintinline{coq}{rewrite} tactics follow the first option, and although we believe there is still interesting research to be done in proof-producing rewriting (see \autoref{sec:prototype:performant:proof-producing-rewriting:theoretical-performance}), our prototype follows the second option.

So, we turn towards the style of using a proven-correct computational procedure within a larger proof engine, known as \emph{proof by reflection}~\cite{ReflectionTACS97}.
Proof by reflection is a well-known strategy for improving performance, and tactics in this style are called \emph{reflective}.
Reflective tactics shine on fixed, well-circumscribed domains.
Coq ships with a number of reflective tactics in the standard library, including solvers for linear and non-linear integer, real, and rational arithmetic~\cite{micromega,romega}; for polynomial positivity over the real field $\mathbb{R}$~\cite{psatz}, for systems of equations over a ring~\cite{ReflectionTACS97}; for polynomial equality over integral domains~\cite{nsatz,gbcoq}.

In contrast, rewriting does not naturally have a fixed domain, such as the structure of rings, fields, integral domains, etc.
%Existing approaches to reflective rewriting tend to be limited in scope or suffer in usability.
\citetfield{aacrewrite} develop a reflective tactic for rewriting modulo associativity and commutativity.
However, the reflective part of the tactic is restricted to just the equations \mintinline{coq}{f x y = f y x} and \mintinline{coq}{f x (f y z) = f (f x y) z}.
  \Rtac{}~\cite{rtac} is a general framework for verified proof tactics in Coq, including an experimental reflective version of \tacrewritestrat{} supporting arbitrary setoid relations, unification variables, and arbitrary semidecidable side conditions solvable by other verified tactics, using de Bruijn indexing to manage binders.

However, \Rtac{} is missing a critical feature for compiling large programs: subterm sharing.
As a result, our experiments with Fiat Cryptography yielded clear asymptotic slowdown.
Furthermore, unlike the convenience of single invocation proof-producing rewriting tactics, \Rtac{} is fairly heavyweight to use!
For instance, \Rtac{} requires that theorems be restated manually in a deep embedding to bring them into automation procedures.
This makes it effort-intensive to adapt \Rtac{} as a rewriting building block to a new project..
%Finally, side conditions of lemmas used in \Rtac{} rewriting must be discharged by fully verified reflective tactics.

\citetfield{Aehlig} come close to a fitting approach, using \emph{normalization by evaluation (NbE)}~\cite{NbE} to bootstrap reduction of open terms on top of full reduction, as built into a proof assistant.\label{sec:aehlig-intro}
However, they expand the proof-assistant trusted code base in ways specific to their technique.
They also do not report any experiments actually using the tool for partial evaluation (just traditional full reduction), potentially hiding performance-scaling challenges or other practical issues.
They also do not preserve subterm sharing explicitly, representing variable references as unary natural numbers (de Bruijn-style).
Finally, they require that rewrite rules be embodied in ML code, rather than stated as natural ``native'' lemmas of the proof assistant.

Thus we find that an adequate proof engine rewriting building block might be built with the following features:

\begin{itemize}
\item Correct: does not extend the trusted code base
\item Complete: handles rewriting on any goal phrasable in the proof assistant
\item Performant: has good asymptotic scaling with input size
\item Convenient: as easy to use as proof-producing rewrite tactics
\item Modular: supports side conditions
\end{itemize}

\section{How Well Does Our Prototype Do?}\label{sec:our-prototype}

The standard of \emph{correctness} we hold ourselves to in developing this building block is \emph{not extending Coq's trusted code base (TCB)}.
By not extending the TCB in any way, our rewriting engine guarantees that building proofs on top of it maintains Coq's rigorous standards for correctness and reliability.
Weaker standards may have use in some applications, but we manage to live up to this exacting standard without too much difficulty.

Our rewriting framework is adequately \emph{performant} into the 100s and 1000s of binders;
 while still a long way from the gigabytes of code required to tackle blue sky projects, this is adequate to handle the real-world demands of Fiat Cryptography.

Much like \Rtac{}, our framework is built around a type of expressions parameterized over an arbitrary enumeration of types and constants.
Extending this modular \emph{convenience} further, we demonstrate how to achieve easy plug-and-play modularity by fully automating the enumerating of type codes, constant codes, and the specialization of our rewriter to these parameters.
Partial evaluation of the reflective rewriting procedure itself nets us additional performance gains.

Similar to \Rtac{} we provide support for decidable side conditions.
However, neither our framework nor \Rtac{} interoperate with the rest of the proof engine as well as necessary for \emph{modularity}.
Similarly,  neither our framework nor \Rtac{}  adequately support dependent\deleted{-}types for \emph{completeness}.
 %While we could have built on top of \Rtac{} to get adequately performant rewriting in Fiat Cryptography, we found the start-up cost too high for the features provided out-of-the-box.

In the context of Fiat Cryptography, there are several critical design criteria and performance bottlenecks that need to be addressed for effective rewriting and partial evaluation, most of which were not provided natively by \Rtac{}.

\begin{itemize}
\item \textbf{Sharing of common subterms:}
  It is essential to represent output programs with shared common subterms for large-scale partial-evaluation problems.
 Inlining shared subterms redundantly can lead to an exponential increase in space requirements.
 The importance of maintaining shared subterms is highlighted in the Fiat Cryptography example of generating a 64-bit implementation of field arithmetic for the P-256 elliptic curve, described in \autoref{sec:under-lets}.

\item \textbf{Proper handling of variable binders:}
  Fiat Cryptography implements the generic arithmetic templates as functional Gallina programs.
  Subterm sharing is \replaced{achieved}{achived} with \mintinline{coq}{let} binders.
  The number of nested variable binders in output terms can be so large that we expect it to be performance-prohibitive to perform bookkeeping operations on first-order-encoded terms (e.g., with de Bruijn indices, as is done in \Rtac{} by \citetfield{rtac}).
  Fiat Cryptography may generate a single routine with nearly a thousand nested binders, emphasizing the need for an efficient rewriting mechanism.
%\todo{Consider being more specific here about expected cost of first-order representation, maybe using some experiment.}
%  \todo{Note from Jason: We haven't actually checked that the bookkeeping is performance-prohibitive.  I suspect there will be a cost, but that if when we made fiat-crypto there was an existing system that used de Bruijn indices and did this, (along with built-in let-lifting, etc), I suspect we might have been a bit more sad, but not prohibitively more sad.  I think we should be much more tentative with this bullet point.}

\item \textbf{Rules with side conditions:}
  Unconditional rewrite rules are generally insufficient, and we require rules with side conditions.
  For instance, Fiat Cryptography depends on checking lack-of-overflow conditions.

\item \textbf{Integration with abstract interpretation:}
  It is not feasible to expect a general engine to discharge all side conditions on the spot.
  We need integration with abstract interpretation that can analyze whole programs to support reduction.

  \item \textbf{Reduction-ordering of rewriting:}
  Existing rewriting tactics order rewrites either in a topdown (starting from the root of the syntax tree) or bottomup (starting from the leaves) manner.
  However, when rewriting is used for doing computation in higher-order functional language, functions play a much more central role than concepts like ``root'' or ``leaves'' of an abstract syntax tree.
  It's much more important to decide whether to evaluate function arguments before or after transforming the body of the function.
  We want to order our rewrites according to something more like call-by-value or call-by-need than topdown or bottomup.
\end{itemize}

\added{\autoref{tab:comparison} summarizes the comparison of our approach with existing approaches to rewriting.}

%Our contributions include answers to a number of challenges that arise in scaling NbE-based partial evaluation in a proof assistant.
Much of the rest of this paper is spent reprising the material from \citetitle{RewritingITP2022}, diving into the details of the design, implementation, and performance evaluation of our rewriting framework.
Readers primarily interested in the high-level insights may wish to skip straight to \autocommanameref{sec:future-work}.

\begin{table}
  \caption{\label{tab:comparison}\added{Comparison of various approaches to rewriting.}}
  \def\tablecheckmark{\textbf{\textcolor{ForestGreen}{\large\checkmark}}}%
  \def\tablecrossmark{\textbf{\textcolor{red}{\large\texttimes}}}%
  \begin{tabular}{p{.47\textwidth}cccc}
    \toprule
    \textbf{Axis of Comparison} & \textbf{Us} & \textbf{\Rtac{}} & \textbf{\usebibentry{Aehlig}{citet}} & \textbf{\mintinline{coq}{rewrite_strat}} \\
    \midrule
    \textbf{Correctness}: Avoids TCB extension & \tablecheckmark & \tablecrossmark\rlap{\footnote{\Rtac{} uses the axiom of functional extensionality.}} & \tablecrossmark\rlap{\footnote{\citetfield{Aehlig} use a trusted compiler.}} & \tablecheckmark \\
    \textbf{Completeness} & \tablecrossmark\footnote{See \autoref{sec:going-further}.} & \tablecrossmark & \tablecheckmark\rlap{\footnote{Completeness is much easier to achieve in a simpler language like HOL.}} & \tablecheckmark \\
    \textbf{Performant} (first-order terms) & \tablecheckmark & \tablecheckmark & \tablecheckmark & \tablecrossmark \\
    \textbf{Performant} (long constant names) & \tablecheckmark & \tablecheckmark & \tablecrossmark\rlap{\footnote{String comparison is used for constant name comparison.}} & \tablecrossmark \\
    \textbf{Performant} (higher-order terms) & \tablecheckmark & ? & ? & \tablecrossmark \\
    \textbf{Performant} (native subterm sharing)\footnote{Native support for common subterm sharing (\autoref{sec:under-lets}) sometimes allows completion of rewriting with asymptotically less term traversal.} & \tablecheckmark & \tablecrossmark & \tablecrossmark & \tablecrossmark \\
    \textbf{Performant} (abstract interpretation) & \tablecrossmark\footnote{We describe how to support integration with abstract interpretation, but we have not yet implemented a fused rewriting and abstract interpretation pass; see \cref{sec:abs-int,sec:future:perf}.} & \tablecrossmark & \tablecrossmark & \tablecrossmark \\
    \textbf{Convenient} & \tablecheckmark & \tablecrossmark\rlap{\footnote{\Rtac{} requires manually writing inductive codes for constants and types and performing semi-manual reification of lemmas.}} & \tablecheckmark & \tablecheckmark \\
    \textbf{Modular}: supports side conditions & $\approx$\rlap{\footnote{Decidable side conditions only}} & $\approx$\rlap{\footnote{Decidable side conditions only}} & ?\rlap{\footnote{\citetfield{Aehlig} seems to indicate that side conditions must be manually added and proven.}} & \tablecheckmark \\
    \textbf{Supports reduction-ordering} & \tablecheckmark & \tablecrossmark & \tablecheckmark & \tablecrossmark \\
    \textbf{Proven correct} (not proof-producing) & \tablecheckmark & \tablecheckmark & \tablecheckmark & \tablecrossmark \\
    \textbf{Avoids explicit binder bookkeeping}\footnote{Explicit binder bookkeeping in the object language might incur performance overhead, though we haven't performed localized measurements.} & \tablecheckmark & \tablecrossmark & \tablecheckmark & \tablecrossmark \\
    \textbf{Proof assistant} & Coq & Coq\rlap{\footnote{Versions 8.5 and 8.6 only}} & \begin{tabular}{c}Isabelle/ \\ HOL \end{tabular} & Coq \\
    \bottomrule
  \end{tabular}
\end{table}

\section{Building a Rewriter}\label{sec:structure}
% POPL 2022 review TODO:
% line 243, "We are [...] but made [...]".
% Stylistically, I suggest sticking with present tense everywhere.
% See also "a type is defined [...] they also wrote [...]" on the next page.
We are mostly guided by \citetfield{Aehlig} but made a number of crucial changes.
Let us review the basic idea of the approach of Aehlig et al.
% POPL 2022 review TODO:
% The description of Aehlig et al.'s approach on page 6 is difficult to follow (for a non-expert reader, at least).
% Why not describe your approach first and contrast it with Aehlig et al.'s approach in the Related Work section?
First, their supporting library contains:
\begin{enumerate}
\item
  Within the logic of the proof assistant (Isabelle/HOL, in their case), a type of syntax trees for ML programs is defined, with an associated (trusted) operational semantics.
\item
  They also wrote a reduction function in (deeply embedded) ML, parameterized on a function to choose the next rewrite, and proved it sound once-and-for-all.
\end{enumerate}

Given a set of rewrite rules and a term to simplify, their main tactic must:
\begin{enumerate}
\item
  \emph{Generate a (deeply embedded) ML program that decides which rewrite rule, if any, to apply at the top node of a syntax tree}, along with a proof of its soundness.
\item
  \emph{Generate a (deeply embedded) ML term standing for the term we set out to simplify}, with a proof that it means the same as the original.
\item
  Combining the general proof of the rewrite engine with proofs generated by reification (the prior two steps), conclude that an application of the reduction function to the reified rules and term is indeed an ML term that generates correct answers.
\item
  ``Throw the ML term over the wall,'' using a general code-generation framework for Isabelle/HOL~\cite{CodeGen}.
  Trusted code compiles the ML code into the concrete syntax of Standard ML, and compiles it, and runs it, asserting an axiom about the outcome.
\end{enumerate}

Here is where our approach differs at that level of detail:
\begin{itemize}
\item
  Our reduction engine is written \emph{as a normal Gallina functional program}, rather than within a deeply embedded language.
  As a result, we are able to prove its type-correctness and termination, and we are able to run it within Coq's kernel.
\item
  We do \emph{compile-time specialization of the reduction engine} to sets of rewrite rules, removing overheads of generality.
\end{itemize}

\subsection{Our Approach in Ten Steps}\label{sec:nine-steps}

Here is a bit more detail on the steps that go into applying our Coq plugin, many of which we expand on in the following sections.
\added{%
Our plugin introduces a new command \mintinline{coq}{Make} for precomputing rewriters from a set of rewrite rules.
For example, we might write:
}
\begin{minted}[fontsize=\small]{coq}
Make myrewriter := Rewriter For (zero_plus, plus_zero, times_zero, times_one).
\end{minted}
\replaced{%
This command automates the following steps:
}{%
For \mintinline{coq}{Make} to precompute a rewriter:
}
\begin{enumerate}
\item
  The given lemma statements are scraped for which named identifiers to encode.
\item
  Inductive types enumerating all available primitive types and functions are emitted.
  This allows us to achieve the performance gains attributed in \citetfield{Boespflug2009} to having native metalanguage constructors for all constants, without manual coding.
\item
  Tactics generate all of the necessary definitions and prove all of the necessary lemmas for dealing with this particular set of inductive codes.
  Definitions include operations like Boolean equality on type codes and lemmas like ``all types have decidable equality.''
\item
  The statements of rewrite rules are reified and soundness and syntactic-well-formedness lemmas are proven about each of them.
  %Each instance of the former involves wrapping the user-provided proof with the right adapter to apply to the reified version.
  %Automating this step allows rewrite rules to be proven in terms of their shallow embedding, which drastically accelerates iteration on the set of rewrite rules.
\item
  Definitions and lemmas needed to prove correctness are assembled into a single package.
\end{enumerate}

Then, to rewrite in a goal\added{, which merely requires invoking a tactic such as \mintinline{coq}{Rewrite_rhs_for myrewriter} or \mintinline{coq}{Rewrite_for myrewriter}}, the following steps are performed \added{automatically}:
\begin{enumerate}
\item
  Rearrange the goal into a single quantifier-free logical formula.%: free-variable quantification is replaced by equality between functions (taking free variables as inputs).
  %\todo{Pedantic Note From Jason: it's not actually an equality between two functions, it's an \mintinline{coq}{equiv} between two functions where \mintinline{coq}{equiv} is a custom relation we define indexed over type codes that is equality up to function extensionality}
\item
  Reify a selected subterm and replace it with a call to our denotation function.
  %We reify the goal side we want to simplify, using inductive codes from the specified package.
  %That goal side is replaced with a denotation-function call on the reified version.
\item
  Rewrite with a theorem, into a form calling our rewriter.
\item Call Coq's built-in full reduction (\tacvmcompute{}) to reduce this application.
\item
  Run standard call-by-value reduction to simplify away use of the denotation function. %on the concrete syntax tree from rewriting.
  %\todo{Is it worth mentioning that we'd instead use \tacvmcompute{} if it supported whitelists (and also said whitelists were recorded in the casts)?}
\end{enumerate}

%We now simultaneously review the approach of \citetfield{Aehlig} and introduce some notable differences in our own approach, noting similarities to the reflective rewriter of \citetfield{rtac} where applicable.

%Let us now describe the language of terms we support rewriting in.
%Note that, while we support rewriting in full-scale Coq proofs, where the metalanguage is dependently typed,
% POPL 2022 review TODO: Line 312: I'm not sure what "nearly simply typed" means.
The object language of our rewriter is nearly simply typed\added{, with limited support for calling polymorphic functions}.
%However, we still support identifiers whose definitions use dependent types, since our reducer does not need to look into definitions.
\begin{align*}
  e ::={}& \phantom{\mid} \texttt{App }e_1\texttt{ }e_2 \mid \texttt{Let }v \defeq e_1\texttt{ In }e_2 %\\
  %&
  \mid \texttt{Abs }(\lambda v.\,e) \mid \texttt{Var }v \mid \texttt{Ident }i
\end{align*}
The \texttt{Ident} case is for identifiers, which are described by an enumeration specific to a use of our library.
For example, the identifiers might be codes for $+$, $\cdot$, and literal constants.
We write $\llbracket e \rrbracket$ for a standard denotational semantics.\label{sec:denote-brackets-def}

\subsection{Pattern-Matching Compilation and Evaluation}\label{sec:pattern-matching-compilation-and-evaluation}

\citetfield{Aehlig} feed a specific set of user-provided rewrite rules to their engine by generating code for an ML function, which takes in deeply embedded term syntax (actually \emph{doubly} deeply embedded, within the syntax of the deeply embedded ML!) and uses ML pattern matching to decide which rule to apply at the top level.
Thus, they delegate efficient implementation of pattern matching to the underlying ML implementation.
As we instead build our rewriter in Coq's logic, we have no such option to defer to ML.
%Indeed, Coq's logic only includes primitive pattern-matching constructs to match one constructor at a time.
%The frontend desugars nested pattern matching into trees of matches, but we do not have access to that desugaring within the language.
%It is arguably a good design decision to keep Coq working this way, since putting efficient nested pattern matching into the kernel would grow the trusted code base substantially.

We could follow a naive strategy of repeatedly matching each subterm against a pattern for every rewrite rule, as in the rewriter of \citetfield{rtac}, but in that case we do a lot of duplicate work when rewrite rules use overlapping function symbols.
Instead, we adopted the approach of \citetfield{maranget2008compiling}, who describes compilation of pattern matches in OCaml to decision trees that eliminate needless repeated work (for example, decomposing an expression into $x + y + z$ only once even if two different rules match on that pattern).
%We have not yet implemented his optimizations for finding \emph{minimal} decision trees.

There are three steps to turn a set of rewrite rules into a functional program that takes in an expression and reduces according to the rules.
The first step is pattern-matching compilation: we must compile the left-hand sides of the rewrite rules to a decision tree that describes how and in what order to decompose the expression, as well as describing which rewrite rules to try at which steps of decomposition.
Because the decision tree is merely a decomposition hint, we require no proofs about it to ensure soundness of our rewriter.
The second step is decision-tree evaluation, during which we decompose the expression as per the decision tree, selecting which rewrite rules to attempt.
The only correctness lemma needed for this stage is that any result it returns is equivalent to picking some rewrite rule and rewriting with it.
The third and final step is to actually rewrite with the chosen rule.
Here the correctness condition is that we must not change the semantics of the expression.
%Said another way, any rewrite-rule replacement expression must match the semantics of the rewrite-rule pattern.

While pattern matching begins with comparing one pattern against one expression, Maranget's approach works with intermediate goals that check multiple patterns against multiple expressions.
A decision tree describes how to match a vector (or list) of patterns against a vector of expressions.
It is built from these constructors:
\begin{itemize}
  \item \texttt{TryLeaf k onfailure}: Try the $k^\text{th}$ rewrite rule; if it fails, keep going with \texttt{onfailure}.
  \item \texttt{Failure}: Abort; nothing left to try.
  \item \texttt{Switch icases app\_case default}:
    With the first element of the vector, match on its kind; if it is an identifier matching something in \texttt{icases}, which is a list of pairs of identifiers and decision trees, remove the first element of the vector and run that decision tree; if it is an application and \texttt{app\_case} is not \texttt{None}, try the \texttt{app\_case} decision tree, replacing the first element of each vector with the two elements of the function and the argument it is applied to; otherwise, do not modify the vectors and use the \texttt{default}.
  \item \texttt{Swap i cont}: Swap the first element of the vector with the $i^\texttt{th}$ element (0-indexed) and keep going with \texttt{cont}.
\end{itemize}

Consider the encoding of two simple example rewrite rules, where we follow Coq's \Ltac{} language in prefacing pattern variables with question marks.
\begin{align*}
  ?n + 0 & \to n %\\
  &
  \texttt{fst}_{\mathbb{Z},\mathbb{Z}}(?x, ?y) & \to x
\end{align*}
We embed them in an AST type for patterns, which largely follows our ASTs for expressions.
\begin{verbatim}
0. App (App (Ident +) Wildcard) (Ident (Literal 0))
1. App (Ident fst) (App (App (Ident pair) Wildcard) Wildcard)
\end{verbatim}
The decision tree produced is\label{sec:compiled-pattern}
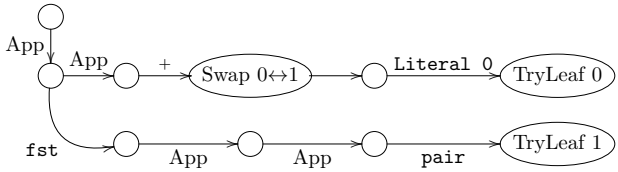
\[\resizebox{230px}{!}{\xymatrix@R-1pc{
  *++[o][F-]\txt{} \ar[d]_{\txt{App}} \\
  *++[o][F-]\txt{} \ar[r]^-{\txt{App}} \ar@/_1.5pc/[dr]_{\txt{\texttt{fst}}} & *++[o][F-]\txt{} \ar[r]^-{+} & *++[o][F-]\txt{Swap 0$\leftrightarrow$1} \ar[r] & *++[o][F-]\txt{} \ar[rr]^-{\txt{\texttt{Literal~0}}} && *++[o][F-]\txt{TryLeaf 0} \\
  & *++[o][F-]\txt{} \ar[r]_-{\txt{App}} & *++[o][F-]\txt{} \ar[r]_-{\txt{App}} & *++[o][F-]\txt{} \ar[rr]_-{\txt{\texttt{pair}}} && *++[o][F-]\txt{TryLeaf 1}
}}\]
\noindent where every nonswap node implicitly has a ``default'' case arrow to \texttt{Failure} and circles represent \texttt{Switch} nodes.

We implement, in Coq's logic, an evaluator for these trees against terms.
Note that we use Coq's normal partial evaluation to turn our general decision-tree evaluator into a specialized matcher to get reasonable efficiency.
Although this partial evaluation of our partial evaluator is subject to the same performance challenges we highlighted in the introduction, it only has to be done once for each set of rewrite rules, and we are targeting cases where the time of per-goal reduction dominates this time of metacompilation.

For our running example of two rules, specializing gives us this match expression.
\label{sec:rewrite-head:example}
\begin{minted}[fontsize=\small]{coq}
match e with
| App f y => match f with
  | Ident fst => match y with
    | App (App (Ident pair) x) y => x | _ => e end
  | App (Ident +) x => match y with
    | Ident (Literal 0) => x | _ => e end | _ => e end | _ => e end.
\end{minted}

\subsection{Adding Higher-Order Features}\label{sec:thunk-eval-subst-term}

Fast rewriting at the top level of a term is the key ingredient for supporting customized algebraic simplification.
However, not only do we want to rewrite throughout the structure of a term, but we also want to integrate with simplification of higher-order terms, in a way where we can prove to Coq that our syntax-simplification function always terminates.
Normalization by evaluation (NbE)~\cite{NbE} is an elegant technique for adding the latter aspect, in a way where we avoid needing to implement our own $\lambda$-term reducer or prove it terminating.

To orient expectations: we would like to enable the following reduction
\begin{align*}
  (\lambda f\ x\ y.\, f\ x\ y)\ (+)\ z\ 0 & \leadsto z
\end{align*}
\noindent using the rewrite rule
\begin{align*}
  ?n + 0 & \to n
\end{align*}

We begin by reviewing NbE's most classic variant, for performing full $\beta$-reduction in a simply typed term in a guaranteed-terminating way.
Our simply typed $\lambda$-calculus syntax is:
\begin{align*}
  t & ::= t \to t ~|~ b
  & e & ::= \lambda v.\, e ~|~ e~e ~|~ v ~|~ c
\end{align*}
\noindent with $v$ for variables, $c$ for constants, and $b$ for base types.

We can now define normalization by evaluation.
First, we choose a ``semantic'' representation for each syntactic type, which serves as an interpreter's result type.
\begin{align*}
  \text{NbE}_{\replaced{T}{t}}(t_1 \to t_2) & \defeq \text{NbE}_{\replaced{T}{t}}(t_1) \to \text{NbE}_{\replaced{T}{t}}(t_2)
  & \text{NbE}_{\replaced{T}{t}}(b) & \defeq \texttt{expr}(b)
\end{align*}
Function types are handled as in a simple denotational semantics, while base types receive the perhaps-counterintuitive treatment that the result of ``executing'' one is a syntactic expression of the same type.
We write $\texttt{expr}(b)$ for the metalanguage type of object-language syntax trees of type $b$, relying on a type family $\texttt{expr}$.

\begin{figure}%[b]
  \small
  \begin{subfigure}[t]{0.5\textwidth}
    \begin{align*}
      \text{reify}_t & : \text{NbE}_{\replaced{T}{t}}(t) \to \text{expr}(t) \\
      \text{reify}_{t_1 \to t_2}(f) & \defeq \lambda v.\,\text{reify}_{t_2}(f(\text{reflect}_{t_1}(v))) \\
      \text{reify}_{b}(f) & \defeq f \\ \noalign{\vskip7pt}
      \text{reflect}_t & : \text{expr}(t) \to \text{NbE}_{\replaced{T}{t}}(t) \\
      \text{reflect}_{t_1\to t_2}(e) & \defeq \lambda x.\,\text{reflect}_{t_2}(e(\text{reify}_{t_1}(x))) \\
      \text{reflect}_{b}(e) & \defeq e \\ \noalign{\vskip7pt}
    \end{align*}
  \end{subfigure}
  \begin{subfigure}[t]{0.5\textwidth}
    \begin{align*}
      \text{reduce} & : \text{expr}(t) \to \text{NbE}_{\replaced{T}{t}}(t) \\
      \text{reduce}(\lambda v. \; e) & \defeq \lambda x. \; \text{reduce}([x/v]e) \\
      \text{reduce}(e_1~e_2) & \defeq \left(\text{reduce}(e_1)\right)(\text{reduce}(e_2)) \\
      \text{reduce}(x) & \defeq x \\
      \text{reduce}(c) & \defeq \text{reflect}(c) \\ \noalign{\vskip7pt}
      \text{NbE} & : \text{expr}(t) \to \text{expr}(t) \\
      \text{NbE}(e) & \defeq \text{reify}(\text{reduce}(e))
    \end{align*}
  \end{subfigure}
\caption{\label{fig:nbe}Implementation of normalization by evaluation}
\end{figure}

Now the core of NbE, shown in \autoref{fig:nbe}, is a pair of dual functions reify and reflect, for converting back and forth between syntax and semantics of the object language, defined by primitive recursion on type syntax.
We split out analysis of term syntax in a separate function reduce, defined by primitive recursion on term syntax\replaced{.
U}{, when u}sually this functionality would be mixed in with reflect\added{, which typically does not need to be recursive over type syntax when constants $c$ can be listed explicitly}.
The reason for this choice will become clear when we extend NbE\added{ to perform rewriting only on fully-applied constants whose $\eta$-long forms do require recursion over type syntax}.

% These definitions apply some of the usual corner-cutting that we see in on-paper descriptions of $\lambda$-term transformations.
We write $v$ for object-language variables and $x$ for metalanguage (Coq) variables, and we overload $\lambda$ notation using the metavariable kind to signal whether we are building a host $\lambda$ or a $\lambda$ syntax tree for the embedded language.
The crucial first clause for reduce replaces object-language variable $v$ with fresh metalanguage variable $x$, and then we are somehow tracking that all free variables in an argument to reduce must have been replaced with metalanguage variables by the time we reach them.
We reveal in \autoref{sec:PHOAS} the encoding decisions that make all the above legitimate, but first let us see how to integrate use of the rewriting operation from the previous section.
To fuse NbE with rewriting, we only modify the constant case of \texttt{reduce}.
% POPL 2022 review TODO: line 433, "First, [...]" but there is no "Second, [...]"?
First, we bind our specialized decision-tree engine (which rewrites \emph{at the root of an AST only}) under the name rewrite-head.

In the constant case, we still reflect the constant, but underneath the binders introduced by full $\eta$-expansion, we perform one instance of rewriting.
In other words, \added{we replace the call to $\text{reflect}(c)$ in the constant case $\text{reduce}(c)$ with a variant of reflect where} we change \replaced{the}{this} one function-definition clause \added{of $\text{reflect}_{b}(e)$ to}:
% POPL 2022 review TODO:
% line 440, is this an amendment to the code in Figure 1?
% It would be clearer to have just one version of the code.
\begin{align*}
  \text{reflect}_{b}(e) & \defeq \text{rewrite-head}(e)
\end{align*}

It is important to note that a constant of function type will be $\eta$-expanded only once for each syntactic occurrence in the starting term, though the expanded function is effectively a thunk, waiting to perform rewriting again each time it is called.
From first principles, it is not clear why such a strategy terminates on all possible input terms.
%, though we work up to convincing Coq of that fact.

The details so far are essentially the same as in the approach of \citetfield{Aehlig}.
Recall that their rewriter was implemented in a deeply embedded ML, while ours is implemented in Coq's logic, which enforces termination of all functions.
Aehlig et al.\ did not prove termination, which indeed does not hold for their rewriter in general, which works with untyped terms, not to mention the possibility of divergent rule-specific ML functions.
In contrast, we need to convince Coq up-front that our interleaved $\lambda$-term normalization and algebraic simplification always terminate.
Additionally, we must prove that rewriting preserves term denotations, which can easily devolve into tedious binder bookkeeping.
%, depending on encoding.

The next section introduces the techniques we use to avoid explicit termination proof or binder bookkeeping, in the context of a more general analysis of scaling challenges.

\section{Scaling Challenges}\label{sec:scaling}

\citetfield{Aehlig} only evaluated their implementation against closed programs.
What happens when we try to apply the approach to partial-evaluation problems that should generate thousands of lines of low-level code?

\subsection{Variable Environments Will Be Large}\label{sec:PHOAS}
We should think carefully about representation of ASTs, since many primitive operations on variables will run in the course of a single partial evaluation.
For instance, \citetfield{Aehlig} reported a significant performance improvement changing variable nodes from using strings to using de Bruijn indices~\cite{debruijn1972}.
However, de Bruijn indices and other first-order representations remain painful to work with.
We often need to fix up indices in a term being substituted in a new context.
Even looking up a variable in an environment tends to incur linear time overhead, thanks to traversal of a list.
Perhaps we can do better with some kind of balanced-tree data structure, but there is a fundamental performance gap versus the arrays that can be used in imperative implementations.
Unfortunately, it is difficult to integrate arrays soundly in a logic.
Also, even ignoring performance overheads, tedious binder bookkeeping complicates proofs.

Our strategy is to use a variable encoding that pushes all first-order bookkeeping off on Coq's kernel or the implementation of the language we extract to, which are themselves performance-tuned with some crucial pieces of imperative code.
Parametric higher-order abstract syntax (PHOAS)~\cite{PhoasICFP08} is a dependently typed encoding of syntax where binders are managed by the enclosing type system.
It allows for relatively easy implementation and proof for NbE, so we adopted it for our framework.

% POPL 2022 review TODO:
% line 504, "Here is the actual inductive definition".
% Could this definition come earlier?
% The definition of reify & reflect would be clearer if it did.
Here is the actual inductive definition of term syntax for our object language, PHOAS-style.
The characteristic oddity is that the core syntax type \texttt{expr} is parameterized on a dependent type family for representing variables.
However, the final representation type \texttt{Expr} uses first-class polymorphism over choices of variable type, bootstrapping on the metalanguage's parametricity to ensure that a syntax tree is agnostic to variable type.
\begin{minted}[fontsize=\small]{coq}
Inductive type := arrow (s d : type) | base (b : base_type).
Infix "→" := arrow.
Inductive expr (var : type -> Type) : type -> Type :=
| Var {t} (v : var t) : expr var t
| Abs {s d} (f : var s -> expr var d) : expr var (s → d)
| App {s d} (f : expr var (s → d)) (x : expr var s) : expr var d
| LetIn {a b} (x : expr var a) (f : var a -> expr var b) : expr var b
| Const {t} (c : const t) : expr var t.
Definition Expr (t : type) : Type := forall var, expr var t.
\end{minted}

The type of base codes \mintinline{coq}{base_type} is constructed automatically based on the types in the specific lemmas the user requests, and might, for example, be
\begin{minted}[fontsize=\small]{coq}
Inductive base_type :=
| Prod (A B : base_type) | List (A : base_type) | Option (A : base_type)
| Unit | Nat | Bool.
\end{minted}

% POPL 2022 review TODO:
% line 517, "A good example of encoding adequacy".
% I believe I know what "adequacy" means in this context, but I don't think every reader will.
A good example of encoding adequacy is assigning a simple denotational semantics.
First, a simple recursive function assigns meanings to types.
\begin{minted}[fontsize=\small]{coq}
Fixpoint denoteT (t : type) : Type := match t with
  | arrow s d => denoteT s -> denoteT d
  | base b    => denote_base_type b  end.
\end{minted}

Next we see the convenience of being able to \emph{use} an expression by choosing how it should represent variables.
Specifically, it is natural to choose \emph{the type-denotation function itself} as variable representation.
% POPL 2022 review TODO:
% line 528, "the suspicious function-abstraction clause".
% I believe I would prefer to see correct code up front.
Especially note how this choice makes rigorous last section's convention (e.g., in the suspicious function-abstraction clause of reduce), where a recursive function enforces that values have always been substituted for variables early enough.
\begin{minted}[fontsize=\small]{coq}
Fixpoint denoteE {t} (e : expr denoteT t) : denoteT t := match e with
  | Var v     => v
  | Abs f     => λ x, denoteE (f x)
  | App f x   => (denoteE f) (denoteE x)
  | LetIn x f => let xv := denoteE x in denoteE f xv
  | Ident c   => denoteI c  end.
Definition DenoteE {t} (E : Expr t) : denoteT t := denoteE (E denoteT).
\end{minted}

It is now easy to follow the same script in making our rewriting-enabled NbE fully formal, in \autoref{fig:nbe2}.
Note especially the first clause of \texttt{reduce}, where we avoid variable substitution precisely because we have chosen to represent variables with normalized semantic values.
The subtlety there is that base-type semantic values are themselves expression syntax trees, which depend on a nested choice of variable representation, which we retain as a parameter throughout these recursive functions.
The final definition $\lambda$-quantifies over that choice.
\added{%
The \mintinline{coq}{rewrite_head} function is automatically generated; the \mintinline{coq}{match} statement in \Vref{sec:rewrite-head:example} is an example of what it might compute.
}

\begin{figure}[b]
  % \begin{subfigure}[t]{0.5\textwidth}
\begin{minted}[fontsize=\small]{coq}
Fixpoint nbeT var (t : type) : Type := match t with
  | arrow s d => nbeT var s -> nbeT var d
  | base b    => expr var b                    end.

Fixpoint reify {var t} : nbeT var t -> expr var t := match t with
  | arrow s d => λ f, Abs (λ x, reify (f (reflect (Var x))))
  | base b    => λ e, e                                      end
with reflect   {var t} : expr var t -> nbeT var t := match t with
  | arrow s d => λ e, λ x, reflect (App e (reify x))
  | base b    => rewrite_head                                end.
Fixpoint reduce {var t} (e : expr (nbeT var) t) : nbeT var t:=match e with
  | Abs e     => λ x, reduce (e (Var x))
  | App e1 e2 => (reduce e1) (reduce e2)
  | Var x     => x
  | Ident c   => reflect (Ident c)                            end.
Definition Rewrite {t} (E : Expr t) : Expr t
  := λ var, reify (reduce (E (nbeT var t))).
\end{minted}
  % \end{subfigure}
  \caption{\label{fig:nbe2}PHOAS implementation of normalization by evaluation}
\end{figure}

One subtlety hidden in \autoref{fig:nbe2} in implicit arguments is in the final clause of \texttt{reduce}, where the two applications of the \texttt{Ident} constructor use different variable representations.
With all those details hashed out, we can prove a pleasingly simple correctness theorem, with a lemma for each main definition, with inductive structure mirroring recursive structure of the definition, also appealing to correctness of last section's pattern-compilation operations.
(We now use syntax $\llbracket \cdot \rrbracket$ for calls to \texttt{DenoteE}.)
$$\forall t, E : \textmintinline{coq}{Expr t}. \; \llbracket \textmintinline{coq}{Rewrite}(E) \rrbracket = \llbracket E \rrbracket$$

% POPL 2022 review TODO:
% line 577, "we needed to convince Coq that the function terminates".
% Coq is convinced already, if it accepts your definition.
% The phrasing is somewhat odd.
%Even before getting to the correctness theorem, we needed to convince Coq that the function terminates.
%While for \citetfield{Aehlig}, a termination proof would have been a whole separate enterprise, it turns out that PHOAS and NbE line up so well that Coq accepts the above code with no additional termination proof; each key function is obviously structurally recursive on either a type or an expression.
% POPL 2022 review TODO:
% line 581, "the Coq kernel is ready to run our Rewrite procedure during checking."
% During checking of what?
% If you mean that the Coq type-checker performs reduction during type-checking, that is a well-known fact which perhaps does not need to be recalled here.
%As a result, the Coq kernel is ready to run our \mintinline{coq}{Rewrite} procedure during checking.

To understand how we now apply the soundness theorem in a tactic, it is important to note how the Coq kernel builds in reduction strategies.
These strategies have, to an extent, been tuned to work well to show equivalence between a simple denotational-semantics application and the semantic value it produces.
In contrast, it is rather difficult to code up one reduction strategy that works well for all partial-evaluation tasks.
Therefore, we should restrict ourselves to (1) running full reduction in the style of functional-language interpreters and (2) running normal reduction on ``known-good'' goals like correctness of evaluation of a denotational semantics on a concrete input.

Operationally, then, we apply our tactic in a goal containing a term $e$ that we want to partially evaluate.
In standard proof-by-reflection style, we \emph{reify} $e$ into some $E$ where $\llbracket E \rrbracket = e$, replacing $e$ accordingly, asking Coq's kernel to validate the equivalence via standard reduction.
Now we use the \mintinline{coq}{Rewrite} correctness theorem to replace $\llbracket E \rrbracket$ with $\llbracket \textmintinline{coq}{Rewrite}(E) \rrbracket$.
Next we ask the Coq kernel to simplify $\textmintinline{coq}{Rewrite}(E)$ by \emph{full reduction via native compilation}.
%since we carefully designed $\mintinline{coq}{Rewrite}(E)$ and its dependencies to produce closed syntax trees, so that reduction will not get stuck pattern-matching on free variables.
Finally, where $E'$ is the result of that reduction, we simplify $\llbracket E' \rrbracket$ with standard reduction.

We have been discussing representation of bound variables.
Also important is representation of constants (e.g., library functions mentioned in rewrite rules).
They could also be given some explicit first-order encoding, but dispatching on, say, strings or numbers for constants would be rather inefficient in our generated code.
Instead, we chose to have our Coq plugin generate a custom inductive type of constant codes, for each rewriter that we ask it to build with \texttt{Make}.
As a result, dispatching on a constant can happen in constant time, based on whatever pattern-matching is built into the execution language (either the Coq kernel or the target language of extraction).
To our knowledge, no past verified reduction tool in a proof assistant has employed that optimization.

%% The payoffs from fully satisfying Coq's type checker are:
%% \begin{enumerate}
%% \item We know that this procedure always terminates, and Coq's kernel is therefore willing to run the procedure for us implicitly during proof checking.  In a sense, we have bootstrapped this reduction strategy into the conversion rule of the type theory.
%% \item In that setting, all bookkeeping about variable binding and environments is handled by the kernel, whose implementation in OCaml allows certain efficient implementation strategies not available to us in the logic.
%% \end{enumerate}

\subsection{Subterm Sharing Is Crucial}\label{sec:under-lets}

For some large-scale partial-evaluation problems, it is important to represent output programs with sharing of common subterms.
Redundantly inlining shared subterms can lead to exponential increase in space requirements.
Consider the Fiat Cryptography~\cite{FiatCryptoSP19} example of generating a 64-bit implementation of field arithmetic for the P-256 elliptic curve.
The library has been converted manually to continuation-passing style, allowing proper generation of \mintinline{coq}{let} binders, whose variables are often mentioned multiple times.
% POPL 2022 review(ish) TODO: actually it's no longer explained further in \autoref{sec:macro}
We ran that old code generator (actually just a subset of its functionality, but optimized by us a bit further, as explained in \autoref{sec:macro}) on the P-256 example and found it took about 15 seconds to finish.
Then we modified reduction to inline \mintinline{coq}{let} binders instead of preserving them, at which point the job terminated with an out-of-memory error, on a machine with 64 GB of RAM.
%(The successful run uses under 2 GB.)
% POPL 2022 review TODO (maybe just include this point):
% Even if the compiler had been able to terminate, the resulting "optimized" code would be practically unusable, as it would perform many redundant computations.

We see a tension here between performance and niceness of library implementation.
When we built the original Fiat Cryptography, we found it necessary to CPS-convert the code to coax Coq into adequate reduction performance.
Then all of our correctness theorems were complicated by reasoning about continuations.
In fact, the CPS reasoning was so painful that at one point most algorithms in the template library were defined twice, once in continuation-passing style and once in direct-style code, because it was easier to prove the two equivalent and work with the non-CPS version than to reason about the CPS version directly.
It feels like a slippery slope on the path to implementing a domain-specific compiler, rather than taking advantage of the pleasing simplicity of partial evaluation on natural functional programs.
Our reduction engine takes shared-subterm preservation seriously while applying to libraries in direct style.

Our approach is \mintinline{coq}{let}-lifting: we lift \mintinline{coq}{let}s to top level, so that applications of functions to \mintinline{coq}{let}s are available for rewriting.
For example, we can perform the rewriting
\begin{align*}
  & \texttt{map}\ (\lambda x.\, y+x)\ (\letin[{z:=e}{[0;1;z+1]}]) \\
  & \; \leadsto \;
  \letin[{z:=e}{[y;y+1;y+(z+1)]}]
\end{align*}
using the rules
\begin{align*}
  \texttt{map}\ {?f}\ [] & \to []
  &
  \texttt{map}\ {?f}\ ({?x} :: {?xs}) & \to f\ x :: \texttt{map}\ f\ xs
  & {?n} + 0 & \to n %\\
\end{align*}

We define a telescope-style type family called \mintinline{coq}{UnderLets}:
\begin{minted}[fontsize=\small]{coq}
Inductive UnderLets {var} (T : Type) := Base (v : T)
| UnderLet {A} (e : @expr var A) (f : var A -> UnderLets T).
\end{minted}
A value of type \mintinline{coq}{UnderLets T} is a series of \texttt{let} binders (where each expression \mintinline{coq}{e} may mention earlier-bound variables) ending in a value of type \mintinline{coq}{T}.
%It is easy to build various ``smart constructors'' working with this type, for instance to construct a function application by lifting the \texttt{let}s of both function and argument to a common top level.

%Such constructors are used to implement an NbE strategy that outputs \mintinline{coq}{UnderLets} telescopes.
Recall that the NbE type interpretation mapped base types to expression syntax trees.
We add flexibility, parameterizing by a Boolean declaring whether to introduce telescopes.

\begin{minted}[fontsize=\small]{coq}
Fixpoint nbeT' {var} (with_lets : bool) (t : type) := match t with
|base t=>if with_lets then @UnderLets var (@expr var t) else @expr var t
|arrow s d=>nbeT' false s -> nbeT' true d  end.
Definition nbeT := nbeT' false.  Definition nbeT_with_lets := nbeT' true.
\end{minted}

%%  - Here are some examples:
%%    - `value Z := UnderLets (expr Z)`
%%    - `value (Z -> Z) := expr Z -> UnderLets (expr Z)`
%%    - `value (Z -> Z -> Z) := expr Z -> expr Z -> UnderLets (expr Z)`
%%    - `value ((Z -> Z) -> Z) := (expr Z -> UnderLets (expr Z)) -> UnderLets (expr Z)`

There are cases where naive preservation of \texttt{let} binders blocks later rewrites from triggering and leads to suboptimal performance, so we include some heuristics.
For instance, when the expression being bound is a constant, we always inline.
When the expression being bound is a series of list ``cons'' operations, we introduce a name for each individual list element, since such a list might be traversed multiple times in different ways.

\subsection{Rules Need Side Conditions}\label{sec:side-conditions}

Many useful algebraic simplifications require side conditions.
For example, bit-shifting operations are faster than divisions, so we might want a rule such as
%One simple case is supporting \emph{nonlinear} patterns, where a pattern variable appears multiple times.
%We can encode nonlinearity on top of linear patterns via side conditions.
\begin{align*}
  {?n} / {?m} & \to n \gg \log_2 m\text{\quad if\quad }2^{\lfloor \log_2 m \rfloor} = m
\end{align*}

The trouble is how to support predictable solving of side conditions during partial evaluation, where we may be rewriting in open terms.
We decided to sidestep this problem by allowing side conditions only as executable Boolean functions, to be applied only to variables that are confirmed as \emph{compile-time constants}, unlike \citetfield{rtac} who support general unification variables.
% POPL 2022 review TODO:
% line 672, "We added a variant of pattern variable [...]".
% This passage is rather obscure.
We added a variant of pattern variable that only matches constants.
Semantically, this variable style has no additional meaning, and in fact we implement it as a special identity function (notated as an apostrophe) that should be called in the right places within Coq lemma statements.
Rather, use of this identity function triggers the right behavior in our tactic code that reifies lemma statements.
%We introduce a notation where a prefixed apostrophe signals a call to the ``constants only'' function.
\label{sec:explain-'}
%We use a special apostrophe marker to indicate a quantified variable that may only match with \emph{compile-time constants}.

Our reification inspects the hypotheses of lemma statements, using type classes to find decidable realizations of the predicates that are used, thereby synthesizing one Boolean expression of our deeply embedded term language, which stands for a decision procedure for the hypotheses.
The \mintinline{coq}{Make} command fails if any such expression contains pattern variables not marked as constants.
%Therefore, matching of rules can safely run side conditions, knowing that Coq's full-reduction engine can determine their truth efficiently.

Hence, we encode the above rule as $\forall n, m. \; 2^{\lfloor \log_2(\texttt{'}m)\rfloor} = \texttt{'}m \to n / \texttt{'}m = n \gg \texttt{'}(\log_2 m)$.

\subsection{Side Conditions Need Abstract Interpretation}\label{sec:abs-int}

With our limitation that side conditions are decided by executable Boolean procedures, we cannot yet handle directly some of the rewrites needed for realistic compilation.
For instance, Fiat Cryptography reduces high-level functional to low-level code that only uses integer types available on the target hardware.
The starting library code works with arbitrary-precision integers, while the generated low-level code should be careful to avoid unintended integer overflow.
As a result, the setup may be too naive for our running example rule ${?n} + 0 \to n$.
When we get to reducing fixed-precision-integer terms, we must be legalistic:
\begin{align*}
  \texttt{add\_with\_carry}_{64}({?n}, 0) & \to (0, n)\text{\ \ if\ \ }0 \le n < 2^{64}
\end{align*}

We developed a design pattern to handle this kind of rule.

First, we introduce a family of functions $\texttt{clip}_{l,u}$, each of which forces its integer argument to respect lower bound $l$ and upper bound $u$.
Partial evaluation is proved with respect to unknown realizations of these functions, only requiring that $\texttt{clip}_{l, u}(n) = n$ when $l \leq n < u$.
Now, before we begin partial evaluation, we can run a verified abstract interpreter to find conservative bounds for each program variable.
When bounds $l$ and $u$ are found for variable $x$, it is sound to replace $x$ with $\texttt{clip}_{l,u}(x)$.
Therefore, at the end of this phase, we assume all variable occurrences have been rewritten in this manner to record their proved bounds.

% POPL 2022 review TODO:
% I believe the authors could perhaps better explain why the rewriting rule at line 706 is preferable (somehow more tractable) than the rewriting rule at line 696.
% At first sight, they seem similar.
% I guess the point is that the variable `n` is unknown (i.e., *not* a compile-time constant) whereas the variable `u` is known (a compile-time constant), so the second rule falls within the scope of the technology described in Section 4.3, whereas the first rule does not.
Second, we proceed with our example rule refactored:
\begin{align*}
  \texttt{add\_with\_carry}_{64}(\texttt{clip}_{\texttt{'}{?l},\texttt{'}{?u}}({?n}), 0) & \to (0, \texttt{clip}_{l,u}(n)) %\\
  %&
  \text{\ \ if\ \ }u < 2^{64}
\end{align*}
If the abstract interpreter did its job, then all lower and upper bounds are constants, and we can execute side conditions straightforwardly during pattern matching.

See \arxivautoreflong{sec:implementation-and-usage} for discussion of some further twists in the implementation.

\subsection{Fusing Compiler Passes}\label{sec:fusing-compiler-passes}

When we moved the
%aforementioned constant-folding rules
%constant-folding rules from \autoref{sec:iteration:carries}
constant-folding rules
from before abstract interpretation to after it, to discharge obligations that we could only prove by bounds analysis, the performance of our compiler on Word-by-Word Montgomery code synthesis decreased significantly.
(The generated code did not change.)
We discovered that the number of variable assignments in our intermediate code was quartic in the number of bits in the prime, while the number of variable assignments in the generated code is only quadratic.
The performance numbers we measured supported this theory: the overall running time of synthesizing code for a prime near $2^k$ jumped from $\Theta(k^2)$ to $\Theta(k^4)$ when we made this change.
We believe that fusing abstract interpretation with rewriting and partial evaluation would allow us to fix this asymptotic-complexity issue.

To make this situation more concrete, consider the following example:
Fiat Cryptography uses abstract interpretation to perform bounds analysis; each expression is associated with a range that describes the lower and upper bounds of values that expression might take on.
Abstract interpretation on addition works as follows: if we have that $x_\ell \le x \le x_u$ and $y_\ell \le y \le y_u$, then we have that $x_\ell + y_\ell \le x + y \le x_u + y_u$.
Performing bounds analysis on $+$ requires two additions.
We might have an arithmetic simplification that says that $x + y = x$ whenever we know that $0 \le y \le 0$.
If we perform the abstract interpretation and then the arithmetic simplification, we perform two additions (for the bounds analysis) and then two comparisons (to test the lower and upper bounds of $y$ for equality with 0).
We cannot perform the arithmetic simplification before abstract interpretation, because we will not know the bounds of $y$.
However, if we perform the arithmetic simplification for each expression after performing bounds analysis on its \emph{subexpressions} and only after this perform abstract interpretation on the resulting expression, then we need not use any additions to compute the bounds of $x + y$ when $0 \le y \le 0$, since the expression will just become $x$.

Another essential pass to fuse with rewriting and partial evaluation is let-lifting.
Unless all of the code is CPS-converted ahead of time, attempting to do let-lifting via rewriting, as must be done when using \tacsetoidrewrite{}, \tacrewritestrat{}, or \Rtac, results in slower asymptotics.
This pattern is already apparent in the example of \Vref{fig:timing-LiftLetsMap}.
%discussed in \autoref{sec:micro:LiftLetsMap}.
\begin{figure}
  {\small %\allowdisplaybreaks
  $\begin{aligned}
    \text{map\_dbl}(\ell) & \defeq \begin{cases} [] & \text{if }\ell = [] \\
        \letin[{y := h + h}{}] & \text{if }\ell = h::t \\
        y :: \text{map\_dbl}(t) &
        \end{cases} \\
    \text{make}(n, m, v) & \defeq \begin{cases} [\underbrace{v, \ldots, v}_n] & \text{if }m = 0 \\
        \text{map\_dbl}(\text{make}(n, m-1, v)) & \text{if }m > 0
        \end{cases} \\
    \text{example}_{n, m} & \defeq \forall v,\ \text{make}(n, m, v) = []
  \end{aligned}$}%
  \caption{\label{fig:micro:LiftLetsMap:code}Initial code for binders and recursive functions}
\end{figure}
% POPL 2022 review TODO:
% line 852, the forward pointer to section 5.2.4 is somewhat unfortunate; the text cannot be fully understood without looking ahead, it seems.
% We do not know what the parameters `n` and `m` represents.
% On line 853, it is not clear what "cubic" means (cubic in what parameter?).
% On line 858, what is `make`?
% What is `map_dbl`? The whole paragraph (848-861) is extremely technical.
% It does not clearly explain why it is the fundamentally difficult to perform let-lifting in an efficient way.
% It also does not explain its opening sentence, "Another essential pass to fuse with rewriting and partial evaluation is let-lifting".
% Why is it essential?
% Is it more efficient if fused than if performed separately?
% Why is that?
Consider the code of \Vref{fig:micro:LiftLetsMap:code}.
We achieve linear performance in $n\cdot m$ when ignoring the final \taccbv,\footnote{%
\added{%
The final \taccbv{} reduces away the denotation function on the AST that results from rewriting.
As discussed in \vref{sentence:final-cbv-perf-blame}, the necessity of ignoring this final reduction step to achieve linear performance is an artifact of measuring performance on a version of Coq prior to 8.14.%
}%
} while \tacsetoidrewrite{} and \tacrewritestrat{} are both cubic.
The rewriter in \Rtac\space cannot possibly achieve better than $\mathcal{O}\left(n\cdot m^2\right)$ unless it can be sublinear in the number of rewrites, because our rewriter gets away with a constant number of rewrites (four), plus evaluating recursion principles for a total amount of work $\mathcal{O}(n\cdot m)$.
But without primitive support for let-lifting, it is instead necessary to lift the lets by rewrite rules, which requires $\mathcal{O}\left(n\cdot m^2\right)$ rewrites just to lift the lets.
The analysis is thus: running \texttt{make} simply gives us $m$ nested applications of \texttt{map\_dbl} to a length-$n$ list.
To reduce a given call to \texttt{map\_dbl}, all existing let-binders must first be lifted (there are $n\cdot k$ of them on the $k$-innermost-call) across \texttt{map\_dbl}, one-at-a-time.
Then the \texttt{map\_dbl} adds another $n$ let binders, so we end up doing $\sum_{k=0}^{m} n\cdot k$ lifts, i.e., $n\cdot m(m+1)/2$ rewrites just to lift the lets.

\section{Evaluation}\label{sec:evaluation}

% TODO: Presumably ``attached to this submission as an anonymized supplement'' will change for the final submission
Our implementation, available on GitHub at \githublink{mit-plv}{rewriter}[ITP-2022-perf-data] and with a roadmap in \arxivautoreflong{sec:CodeSupplement-more}, includes a mix of Coq code for the proved core of rewriting, tactic code for setting up proper use of that core, and OCaml plugin code for the manipulations beyond the tactic language's current capabilities.
We report here on evidence that the tool is effective, first in terms of productivity by users and then in terms of compile-time performance.

\subsection{Iteration on the Fiat Cryptography Compiler}\label{sec:iteration}

We ported Fiat Cryptography's core compiler functionality to use our framework.
The result is now used in production by a number of open-source projects.
We were glad to retire the CPS versions of verified arithmetic functions, which had been present only to support predictable reduction with subterm sharing.
More importantly, it became easy to experiment with new transformations via proving new rewrite theorems, directly in normal Coq syntax, including the following, all justified by demand from real users:
\begin{itemize}
\item Reassociating arithmetic to minimize the bitwidths of intermediate results
\item Multiplication primitives that separately return high halves and low halves
\item Strings and a ``comment'' function of type $\forall A. \; \texttt{string} \to A \to A$
\item Support for bitwise exclusive-or
\item A special marker to block C compilers from introducing conditional jumps in code that should be constant-time
\item Eliding bitmask-with-constant operations that can be proved as no-ops
\item Rules to introduce conditional moves (on supported platforms)
\item New hardware backend, via rules that invoke special instructions of a cryptographic accelerator
\item New hardware backend, with a requirement that all intermediate integers have the same bitwidth, via rules to break wider operations down into several narrower operations
\end{itemize}

% This is an odd one to include in my summarized list.
% It sounds like more of an observation on the software-engineering experience,
% rather than extending fiat-crypto.  Maybe BRING BACK if we get more space.
%% \subsubsection{Moving Rules Involving Carries}\label{sec:iteration:carries}
%% We originally had rules like ``adding 0 to $x$ produces just $x$, with no carry.''
%% This rule is not true in general, because we must encode carrying as happening at a particular bitwidth, while we have no guarantee that $x$ fits within that bitwidth.
%% The correct rule has a precondition: that $x$ is between 0 and $2^{64}$.

%% We discovered this issue when trying to prove our rewrite rules correct.
%% As a result, we had to move this sort of rule from happening before abstract interpretation to happening after abstract interpretation.
%% Since the passes are just defined as lists of rewrite rules, moving the rules was quite simple.
%(Rephrasing them to talk about the outputs of abstract interpretation was somewhat painful, though, because the proof assistant did not enforce the conventions we were using for where to store abstract-interpretation information.
%We hypothesize that a more uniform approach to integrating abstract interpretation with rewriting and partial evaluation would solve this problem.)

%% BRING BACK \label{sec:fusing-compiler-passes} subsubsection from appendix when we have plenty of space.

\subsection{Microbenchmarks}\label{sec:micro}

\gdef\NoBindersSubfloatNval{3}%
\def\NoBindersSubfloatXRow{\thisrow{param-2-m}*(2^(\thisrow{param-1-n}+1)-1)}%

\begin{figure*}
  \newsavebox{\NestedBindersSubfloat}%
  \sbox{\NestedBindersSubfloat}{%
    \adjustbox{valign=t}{\resizebox{0.32\textwidth}{!}{\beginTikzpictureStamped[only marks]{
      \input{perf-UnderLetsPlus0.csv.md5} \space
    }
      \pgfplotsset{every axis legend/.append style={
          at={(0.5,-0.2)},
          anchor=north}}
      \begin{axis}[xlabel=\# of let binders,
          ylabel=time (s),
          scaled x ticks=false,
          ymax=65,
          xmax=5000,
          table/col sep=comma,
          table/x=param-n
        ]
        \addplot[mark=o,color=red]         table[y=rewrite-strat(bottomup)-regression-exponential-user]{perf-UnderLetsPlus0.csv};
        \addplot[mark=triangle,color=red]  table[y=rewrite-strat(topdown)-regression-exponential-user] {perf-UnderLetsPlus0.csv};
        \addplot[mark=square,color=red]    table[y=setoid-rewrite-regression-cubic-user]               {perf-UnderLetsPlus0.csv};
        \addplot[mark=+,color=blue]        table[y=Rewrite-for-gen-user]                               {perf-UnderLetsPlus0.csv};
        \addplot[mark=x,color=ForestGreen] table[y=rewriting-user]                                     {perf-UnderLetsPlus0.csv};
        \legend{rewrite\_strat bottomup,rewrite\_strat topdown,setoid\_rewrite,{Our approach including reification, cbv, etc.},Our approach (only rewriting)}
      \end{axis}
    \end{tikzpicture}}}}%
  \newsavebox{\BindersAndRecursiveFunctionsSubfloat}%
  \sbox{\BindersAndRecursiveFunctionsSubfloat}{%
    \adjustbox{valign=t}{\resizebox{0.32\textwidth}{!}{\beginTikzpictureStamped[only marks]{
        \input{perf-UnderLetsPlus0.csv.md5} \space
    }
      \pgfplotsset{every axis legend/.append style={
          at={(0.5,-0.2)},
          anchor=north}}
      \begin{axis}[xlabel={$m \cdot n$ (${}={}$\# of let binders${}={}$\# of recursive calls)},
          ylabel=time (s),
          scaled x ticks=false,
          ymax=27,
          xmax=12000,
          table/col sep=comma,
          table/x=param-0-nm]
        \addplot[mark=o,color=red]         table[y=rewrite-strat(bottomup-bottomup)-regression-exponential-user]{perf-LiftLetsMap.csv};
        \addplot[mark=triangle,color=red]  table[y=rewrite-strat(topdown-bottomup)-regression-exponential-user]{perf-LiftLetsMap.csv};
        \addplot[mark=square,color=red]    table[y=setoid-rewrite-regression-cubic-user]{perf-LiftLetsMap.csv};
        \addplot[mark=+,color=blue]        table[y=Rewrite-for-gen-user]{perf-LiftLetsMap.csv};
        \addplot[mark=x,color=ForestGreen] table[y=rewriting-user]{perf-LiftLetsMap.csv};
        \addplot[mark=*,color=purple]      table[y=cps+vm-compute-regression-quadratic-user]{perf-LiftLetsMap.csv};
        \legend{rewrite\_strat bottomup,rewrite\_strat topdown,repeat setoid\_rewrite,{Our approach including reification, cbv, etc.},Our approach (only rewriting),cps+vm\_compute}
      \end{axis}
    \end{tikzpicture}}}}%
  \newsavebox{\NoBindersSubfloat}%
  \sbox{\NoBindersSubfloat}{%
    \edef\nval{\NoBindersSubfloatNval}%
    \adjustbox{valign=t}{\resizebox{0.32\textwidth}{!}{\beginTikzpictureStamped[only marks]{
        \input{perf-Plus0Tree.csv.md5} \space
        \nval
    }
      \pgfplotsset{every axis legend/.append style={
          at={(0.5,-0.2)},
          anchor=north}}
      % since n = 1, we have 2^n=twice as many rewrite locations as the value on the x axis, so we need to double things
      %,xticklabel={\pgfkeys{/pgf/fpu}\pgfmathparse{2^\nval*\tick}$\mathsf{\pgfmathprintnumber{\pgfmathresult}}$}
      \begin{axis}[xlabel=\# of rewrite locations,
          scaled x ticks=false,
          ylabel=time (s),
          ymax=7,
          xmax=15000,
          xtick distance=3000,
          table/col sep=comma,
          table/x expr={\NoBindersSubfloatXRow}]% ymax=10]
        \addplot[discard if not={param-1-n}{\nval},mark=square,color=red] table[y=rewrite-strat(bottomup)-regression-cubic-user]{perf-Plus0Tree.csv};
        \addplot[discard if not={param-1-n}{\nval},mark=*,color=red] table[y=setoid-rewrite-regression-cubic-user]{perf-Plus0Tree.csv};
        \addplot[discard if not={param-1-n}{\nval},mark=triangle,color=red] table[y=rewrite-strat(topdown)-regression-cubic-user]{perf-Plus0Tree.csv};
        \addplot[discard if not={param-1-n}{\nval},mark=o,color=red] table[y=rewrite!-regression-cubic-user]{perf-Plus0Tree.csv};
        \addplot[discard if not={param-1-n}{\nval},mark=diamond,color=red] table[y=ssr-rewrite!-regression-cubic-user]{perf-Plus0Tree.csv};
        \addplot[discard if not={param-1-n}{\nval},mark=+,color=blue] table[y=Rewrite-for-gen-user]{perf-Plus0Tree.csv};
        \addplot[discard if not={param-1-n}{\nval},mark=x,color=ForestGreen] table[y=rewriting-user]{perf-Plus0Tree.csv};
        \legend{rewrite\_strat bottomup,setoid\_rewrite,rewrite\_strat topdown,rewrite!,ssreflect rewrite!,{Our approach including reification, cbv, etc.},Our approach (only rewriting)}
      \end{axis}
    \end{tikzpicture}}}}%
  \newsavebox{\FiatCryptoSubfloat}%
  \sbox{\FiatCryptoSubfloat}{%
    \adjustbox{valign=t}{\resizebox{0.32\textwidth}{!}{\beginTikzpictureStamped[only marks]{
      \input{perf-fiat-crypto.csv.md5} \space
%      \einput{perf-fiat-crypto-rewriting.csv.md5}
    }
      \pgfplotsset{every axis legend/.append style={
          at={(0.5,-0.2)},
          anchor=north}}
      \begin{axis}[xlabel=prime,ylabel=time relative to original Fiat Crypto,xmode=log, ymode=log,log basis x={2},
%          extra x ticks={2^255,2^256,2^384},
%          extra x tick labels={\formatextratick{Curve25519},\formatextratickLower{P-256},\formatextratick{P-384}},
%          extra x tick label style={yshift={-1em}},
        ]
        \addplot[mark=*,color=red] table[x=prime,y=OldSynthesisAndPackage-over-OldSynthesisAndPackage-real,col sep=comma]{perf-fiat-crypto.csv};
        \addlegendentry{Original Fiat Crypto (includes reification+rewriting)}
        \addplot[mark=+,color=blue] table[x=prime,y=NewVMFull-over-OldSynthesisAndPackage-real,col sep=comma]{perf-fiat-crypto.csv};
        \addlegendentry{Our approach w/ Coq's VM}

        %\addplot[mark=o,color=purple] table[x=prime,y=OldVM-over-OldSynthesisAndPackage-real,col sep=comma]{perf-fiat-crypto.csv};
        %\addlegendentry{Old approach (handwritten-CPS+VM, only reduction)}

        \addplot[mark=x,color=ForestGreen] table[x=prime,y=NewExtractionFull-over-OldSynthesisAndPackage-real,col sep=comma]{perf-fiat-crypto.csv};
        \addlegendentry{Our approach w/ extracted OCaml}
      \end{axis}
    \end{tikzpicture}}}}%
  \newsavebox{\FiatCryptoSetoidSubfloat}%
  \sbox{\FiatCryptoSetoidSubfloat}{%
    \adjustbox{valign=t}{\resizebox{0.32\textwidth}{!}{\beginTikzpictureStamped[only marks]{
      \input{perf-fiat-crypto.csv.md5} \space
      \input{perf-fiat-crypto-rewriting.csv.md5} \space
    }
      \pgfplotsset{every axis legend/.append style={
          at={(0.5,-0.2)},
          anchor=north}}
      \begin{axis}[xlabel=prime,ylabel=time (s),xmode=log, ymode=log,log basis x={2},
%          extra x ticks={2^255,2^256,2^384},
%          extra x tick labels={\formatextratick{Curve25519},\formatextratickLower{P-256},\formatextratick{P-384}},
%          extra x tick label style={yshift={-1em}},
        ]
        \addplot[mark=o,color=purple] table[x=prime,y=partial-eval-and-rewrite-x64-time,col sep=comma]{perf-fiat-crypto-rewriting.csv};
        \addlegendentry{Proof-producing rewriting + partial evaluation}
        \addplot[mark=*,color=red] table[x=prime,y=UnsaturatedSolinas-x64-OldSynthesisAndPackage-real,col sep=comma]{perf-fiat-crypto.csv};
        \addlegendentry{Original Fiat Crypto (including reification+rewriting)}
        \addplot[mark=+,color=blue] table[x=prime,y=UnsaturatedSolinas-x64-NewVMFull-real,col sep=comma]{perf-fiat-crypto.csv};
        \addlegendentry{Our approach w/ Coq's VM}
        \addplot[mark=x,color=ForestGreen] table[x=prime,y=UnsaturatedSolinas-x64-NewExtractionFull-real,col sep=comma]{perf-fiat-crypto.csv};
        \addlegendentry{Our approach w/ extracted OCaml}
      \end{axis}
    \end{tikzpicture}}}}%
  \newcommand{\vphantomSubfloatOne}{%
    \vphantom{\usebox{\NestedBindersSubfloat}}%
    \vphantom{\usebox{\NoBindersSubfloat}}%
    \vphantom{\usebox{\BindersAndRecursiveFunctionsSubfloat}}%
  }%
  \newcommand{\vphantomSubfloatTwo}{%
    \vphantom{\usebox{\FiatCryptoSetoidSubfloat}}%
    \vphantom{\usebox{\FiatCryptoSubfloat}}%
  }%
  \centering
  \subfloat[No binders]{\usebox{\NoBindersSubfloat}\vphantomSubfloatOne\label{fig:timing-Plus0Tree}}%
  %\quad
  \subfloat[Nested binders]{\usebox{\NestedBindersSubfloat}\vphantomSubfloatOne\label{fig:timing-UnderLetsPlus0}}%
  %\quad
  \subfloat[\centering Binders and recursive functions]{\usebox{\BindersAndRecursiveFunctionsSubfloat}\vphantomSubfloatOne\label{fig:timing-LiftLetsMap}}%
  \\
  %\quad
  \subfloat[\centering Fiat Cryptography \mbox{(relative timing)}]{\usebox{\FiatCryptoSubfloat}\vphantomSubfloatTwo\label{fig:fiat-crypto-scaling}}
  \quad
  \subfloat[\centering Fiat Cryptography (absolute timing, only unsaturated Solinas x64)]{\usebox{\FiatCryptoSetoidSubfloat}\vphantomSubfloatTwo\label{fig:fiat-crypto-scaling-with-setoid}}

  \caption{Timing of different partial-evaluation implementations}\label{fig:multi-timing}
\end{figure*}

% POPL 2022 review TODO:
% The graphs in Figure 3 are pretty ugly because they have so many data points packed together.
% It is nonetheless possible to distinguish the various series involved though, with a bit of geometrical deduction.
% And I suppose you'd lose some information if you thinned out the data points or plotted lines only.
% So I don't really have a concrete suggestion for improving the graphs; I just don't find them very pleasing to look at.

Now we turn to evaluating performance of generated compilers.
We start with microbenchmarks focusing attention on particular aspects of reduction and rewriting, with \arxivautoreflong{sec:additionalMicro} going into more detail, including on a few more benchmarks.

Our first example family, \emph{nested binders}, has two integer parameters $n$ and $m$.
An expression tree is built with $2^n$ copies of an expression, which is itself a free variable with $m$ ``useless'' additions of zero.
We want to see all copies of this expression reduced to just the variable.
\Vref{fig:timing-Plus0Tree} shows the results for $n = \NoBindersSubfloatNval$ as we scale $m$.
The comparison points are Coq's \texttt{rewrite!}, \tacsetoidrewrite{}, and \tacrewritestrat{}.
The first two perform one rewrite at a time, taking minimal advantage of commonalities across them and thus generating quite large, redundant proof terms for large, redundant statements.
The third makes top-down or bottom-up passes with combined generation of proof terms and claims.
For our own approach, we list both the total time and the time taken for core execution of a verified rewrite engine, without counting reification (converting goals to ASTs) or its inverse (interpreting results back to normal-looking goals).
The comparison here is very favorable for our approach so long as $m > 2$.
%% The competing tactics spike upward toward timeouts at just around a thousand rewrite locations, while our engine is still under two seconds for examples with tens of thousands of rewrite locations.
%% When $m < 2$, Coq's \texttt{rewrite!} tactic does a little bit better than our engine, corresponding roughly to the overhead incurred by our term representation (which, for example, stores the types at every application node) when most of the term is in fact unchanged by rewriting.
(See \arxivautoreflong{sec:additionalPlots:Plus0Tree} %\footnote{Like several forward references in this section, this one goes to an appendix included within the main submission page limit, to avoid interrupting the flow in presenting the most important results.}
for more detailed plots.)

Now consider what happens when we use \mintinline{coq}{let} binders to share subterms within repeated addition of zero, incorporating exponentially many additions with linearly sized terms.
\Vref{fig:timing-UnderLetsPlus0} shows the results.
The comparison here is again very favorable for our approach.
The competing tactics spike upward toward timeouts at just a few hundred generated binders, while our engine is only taking about 10 seconds for examples with 5,000 nested binders.

When recursive functions are mixed with \mintinline{coq}{let} binders, we must lift the \mintinline{coq}{let} binders across the functions to avoid blocking reduction and expose more rewriting opportunities.
We may either lift \mintinline{coq}{let} binders with additional equations passed to \tacsetoidrewrite{} or \tacrewritestrat{}, build in support for let-lifting (our approach), or manually rewrite the code in continuation passing style and module-opacify the constants which are not to be unfolded (to take advantage of Coq's built-in VM reduction without our rewriter).
Note that this last option is available for this example because it only involves partial reduction and not equational rewriting.
\Vref{fig:timing-LiftLetsMap} shows the results of our evaluation on the code in \Vref{fig:micro:LiftLetsMap:code}.
The prior state of the art---writing code in CPS---suitably tweaked by using module opacity to allow \tacvmcompute{}, remains the best performer here, though the cost of rewriting everything is CPS may be prohibitive.
Our method soundly beats \tacrewritestrat{}.
\replaced{When we collected the data for \autoref{fig:multi-timing}, we were}{We are} additionally bottlenecked on \taccbv{}, which is used to unfold the goal post-rewriting and cost\deleted{s} about a minute on the largest of terms; about 99\% of the difference between the full time of our method and just the rewriting is spent in the final \taccbv{} at the end, used to denote our output term from reified syntax.
\label{sentence:final-cbv-perf-blame}%
\replaced{This}{We blame this} performance \replaced{bottleneck is an artifact of}{on} the unfortunate fact that reduction in Coq \replaced{was}{is} quadratic in the number of nested binders present\added{ when we collected our performance data}; see \coqbug{11151}.
This bug has since been fixed, as of Coq 8.14; see \coqpr{13537}.

%Additionally, we consider another option, which was adopted by Fiat Cryptography at a larger scale: rewrite our functions to improve reduction behavior.
% POPL 2022 review TODO:
% line 958 (and elsewhere), "continuation-passing style [...] allows standard VM-based reduction to achieve good performance".
% Is it folklore knowledge why writing in CPS style helps?
% It might be worth briefly recalling why this is so.
%Specifically, both functions are rewritten in continuation-passing style, which makes them harder to read and reason about but allows standard VM-based reduction to achieve good performance.
%The figure shows that \tacrewritestrat{} variants are essentially unusable for this example, with \tacsetoidrewrite{} performing only marginally better, while our approach applied to the original, more readable definitions loses ground steadily to VM-based reduction on CPS'd code.
%On the largest terms ($n \cdot m > 20,000$), the gap is 6s vs.\ 0.1s of compilation time, which should often be acceptable in return for simplified coding and proofs, plus the ability to mix proved rewrite rules with built-in reductions.

%traded for simplified coding and proofs, plus the ability to mix proved rewrite rules with built-in reductions.

Although we have made our comparison against the built-in tactics \tacsetoidrewrite{} and \tacrewritestrat{}, by analyzing the performance in detail, we can argue that these performance bottlenecks are likely to hold for any proof assistant designed like Coq.
Detailed debugging reveals six performance bottlenecks in the existing tactics, already discussed in \autoref{sec:setoid-rewrite-bottlenecks}.

\subsection{Macrobenchmark: Fiat Cryptography}\label{sec:macro}

Finally, we consider an experiment (described in more detail in \arxivautoreflong{sec:additionalMacro}) replicating the generation of performance-competitive finite-field-arithmetic code for all popular elliptic curves by \citetfield{FiatCryptoSP19}.
In all cases, we generate essentially the same code as they did, so we only measure performance of the code-generation process.
We stage partial evaluation with three different reduction engines (i.e., three \mintinline{coq}{Make} invocations), respectively applying 85, 56, and 44 rewrite rules (with only 2 rules shared across engines), taking total time of about 5 minutes to generate all three engines.
These engines support 95 distinct function symbols.

\Vref{fig:fiat-crypto-scaling} graphs running time of three different partial-evaluation and rewriting methods for Fiat Cryptography, as the prime modulus of arithmetic scales up.
Times are normalized to the performance of the original method of \citetfield{FiatCryptoSP19}, which relied on standard Coq reduction to evaluate code that had been manually written in CPS, followed by reification and a custom ad-hoc simplification and rewriting engine.
%In the course of running this experiment, we found a way to improve the reduction part of the old approach for a somewhat fairer comparison.
%It had relied on Coq's configurable \taccbv{} tactic to perform reduction with selected rules of the definitional equality, which the Fiat Cryptography developers had applied to blacklist identifiers that should be left for compile-time execution.
%By instead hiding those identifiers behind opaque module-signature ascription, we were able to run Coq's more-optimized virtual-machine-based reducer.

%As the figure shows, our approach running partial evaluation inside Coq's kernel begins with about a 10$\times$ performance disadvantage vs.\ the original method.
%With log scale on both axes, we see that this disadvantage narrows to become nearly negligible for the largest primes, of around 500 bits.
As the figure shows, our approach gives about a 10$\times$--1000$\times$ speed-up over the original Fiat Cryptography pipeline.
%(We used the same set of prime moduli as in the experiments run by \citetfield{FiatCryptoSP19}, which were chosen based on searching the archives of an elliptic-curves mailing list for all prime numbers.)
Inspection of the timing profiles of the original pipeline reveals that reification dominates the timing profile; since partial evaluation is performed by Coq's kernel, reification must happen \emph{after} partial evaluation, and hence the size of the term being reified grows with the size of the output code.
Also recall that the old approach required rewriting Fiat Cryptography's library of arithmetic functions in continuation-passing style, enduring this complexity in library correctness proofs, while our new approach applies to a direct-style library.
Finally, the old approach included a custom reflection-based arithmetic simplifier for term syntax, run after traditional reduction, whereas now we are able to apply a generic engine that combines both, without requiring more than proving traditional rewrites.

The figure also confirms a clear performance advantage of running reduction in code extracted to OCaml, which is possible because our plugin produces verified code in Coq's functional language.
The extracted version is about 10$\times$ faster than running in Coq's kernel.
%By the time we reach middle-of-the-pack prime size around 300 bits, the extracted version, which includes arithmetic simplification, is faster even than the best performance we could achieve for the old

\Vref{fig:fiat-crypto-scaling-with-setoid} graphs running time of the same three partial-evaluation and rewriting methods for Fiat Cryptography, in addition to the impractical \tacrewritestrat{}-based method, as the prime modulus of arithmetic scales up.

\section{Future Work}\label{sec:future-work}

By far the biggest next step for our engine is to integrate abstract interpretation with rewriting and partial evaluation.
We expect this would net us asymptotic performance gains as described in \autoref{sec:fusing-compiler-passes}.
Additionally, it would allow us to simplify the phrasing of many of our post-abstract-interpretation rewrite rules, by relegating bounds information to side conditions rather than requiring that they appear in the syntactic form of the rule.

There are also a number of natural extensions to our engine.
For instance, we do not yet allow pattern variables marked as ``constants only'' to apply to container datatypes; we limit the mixing of higher-order and polymorphic types, as well as limiting use of first-class polymorphism; we do not support rewriting with equalities of nonfully-applied functions; we only support decidable predicates as rule side conditions, and the predicates may only mention pattern variables restricted to matching constants; we have hardcoded support for a small set of container types and their eliminators; we support rewriting with equality and no other relations% (e.g., subset inclusion)
; and we require decidable equality for all types mentioned in rules.
%It may be helpful to design an engine that lifts some or all of these limitations, building on the basic structure that we present here.

\subsection{What Would It Take For Our Prototype To Be A Full-Fledged Proof Engine Building Block?}\label{sec:going-further}

We return now to the context of our introduction: having a rewriting proof engine building block that performs adequately at scale.

\subsubsection{Performance}\label{sec:future:perf}
While we achieve adequate performance on the real-world demands of Fiat Cryptography, there is much work left to be done.
While we can handle 100s--1\,000s of lines of code in a single function, easily accomodating even the largest of commonly-used primes in ECC, other uses of finite field arithmetic use much larger primes.
For example, the uses for Bitcoin involve primes such as $2^{3072} - 1103717$.
At over 3000 bits and 48 limbs, we'd need to be able to handle generating nearly 42\,000 lines of code.
% 40.4+12.2*48+17.9*48^2 = 41867.6
(In fact the situation is worse: unless we fuse rewriting with abstract interpretation, our intermediate code will be nearly 11.5 million lines long.\footnote{%
See \href{https://github.com/mit-plv/fiat-crypto/issues/851\#issuecomment-662710320}{Fiat Crypto Issue \#851: Support for large finite fields} for more details.%
})
% 3.57+10*48+2.57*48^2+8.26*48^3+1.98*48^4 = 11430558.45
At this rate, it would take approximately somewhere between a week and a month to generate the code for this prime.\footnote{%
Plotting the time of code generation for our extracted tool on WBW Montgomery on x32 and x64 as a function of computed intermediate lines of code $\ell$ (for $n$ limbs, $\ell \approx 3.57+10n+2.57n^2+8.26n^3+1.98n^4$) shows a definite superlinear trend.
The best-fit quadratic ($R^2 > 0.995$) is $\text{\# seconds} = 0.803 + 1.94\cdot 10^{-4}\ell + 4.28\cdot 10^{-9}\ell^2$ for x32 and $\text{\# seconds} = 3.42\cdot 10^{-5} + 2.84\cdot 10^{-4}\ell + 1.6\cdot 10^{-8}\ell^2$ for x64.
For 48 limbs, this comes out to 2\,093\,768s ($\approx$ 6.5 days) or 561\,433s ($\approx$ 24 days), respectively.%
}

Achieving adequate performance on code that is gigabytes or terabytes in size is an open research question!

\subsubsection{Modularity in Side Conditions}\label{sec:future:modular}
As discussed at the end of Related \& Future Work in \citechapter{rtac}, reflective procedures cannot invoke unverified tactic automation.
\citetfield{rtac} suggest ``native support for invoking external procedures and reconstructing the results in Coq \emph{a la} Claret's work~\cite{claret2013lightweight}.''
Another possibility might be to make Coq's efficient computation routines reentrant:
when an existential variable or other special marker shows up during reduction, the reduction tactic might be able to pop back into interactive tactic mode, allowing the user to partially fill the existential variable with other tactics before resuming efficient computation.

\subsubsection{Scaling Up the Scope}\label{sec:future:complete}

The final deficiency of our prototype rewriting tactic is a lack of support for the full scope of Coq's mathematical language.
We cannot handle most dependent types, bare (co)fixpoint constructs, primitive integers and floats, etc.

Hence we conclude this paper by laying out a research agenda for scaling up our prototype rewriting tactic to be a fully adequate replacement for the built-in rewriting tactics.

The first step is to upgrade the term representation to something like MetaCoq's~\cite{metacoq,malecha2015thesis,coq-coq-correct} AST, which can faithfully represent all terms accepted by the Coq kernel.

Five obstacles remain to writing a denotation function, which is essential for making use of reflective automation.
We posit that all five obstacles can be overcome with the same kind of automatic specialization automation we use in our prototype to allow easy rewriting on supported domains.

\paragraph{The G\"odelian Obstacle}
G\"odel's incompleteness theorem~\cite{godel1931incompleteness} says that no consistent system can prove its own consistency.
Even more directly, L\"ob's theorem~\cite{lob1955theorem} tells us that any total denotation function---a necessary part of reflective automation---gives rise to a proof of \mintinline{coq}{False}.%
\footnote{%
\added{%
L\"ob's theorem says $(\square P \to P) \to P$ for any proposition $P$.%
}%
\footnotemark{}
\added{%
We may interpret $\square P$ as ``a proof of $P$'' or, by Curry-Howard, ``an abstract syntax tree for a term of type $P$''.
Functions of type $\square P \to P$ \emph{are just} total denotation functions from syntax to semantics: they consume abstract syntax trees well-typed at $P$ and produce inhabitants of $P$.
Instantiating $P$ with \mintinline{coq}{False} completes the claim.%
}%
}
\footnotetext{%
\added{%
Other variants include $\square(\square P \to P) \to \square P$, $\square(\square(\square P \to P) \to \square P)$, and $(\vdash (\square P \to P)) \Longrightarrow (\vdash P)$.
The assumptions used to prove each variant differ only slightly, and we elide them in this footnote.
}%
}

Folklore has it that strong normalization of the theory of Coq with $n$ universes can be proven in Coq with $n+k$ universes for some $0 < k \le 4$~\cite{pujet2023logrelmltt,westbrook2011uniform}.

By parameterizing the rewriter over an arbitrary universe graph and autospecializing to a desired universe graph on-the-fly, we could in theory handle any number of universes, thus bypassing the G\"odelian obstacle.

\paragraph{Named Constants and Inductives}
Just as in our prototype, the denotation function will need to be paraterized over a mapping of named constants, (co)inductive types, constructors, and eliminators.
The same sort of autospecialization that our prototype uses should handle the full scope of types and constants available in Coq without any problem.
For performance, the term representation should perhaps be parameterized over named constants and (co)inductives, though, rather than using strings as is currently done in MetaCoq.

\paragraph{(Co)Fixpoints, Case Analysis, and the Guard Condition}
Unlike Lean, Coq has primitive constructs for general case analysis and guarded (co)recursion.
We cannot translate these constructs directly, because the \mintinline{coq}{fix}, \mintinline{coq}{cofix}, and \mintinline{coq}{match} constructs are not first-class terms in Coq.
However, autospecialization can rescue us again, by generating on-the-fly the set of anonymous recursive functions and case analyses that are present in the term we are rewriting in and the lemmas we use for rewriting.

\paragraph{Judgmental Equality and Reduction: The Semisimplicial Obstacle}
Even if we manage to automatically generate strong normalization proofs for our rewriting building block, there remains one final (forseeable) obstacle to writing a total denotation function in an intensional type theory: judgmental equality.
This problem can be most easily seen by considering the semisimplicial types.
This is an infinite telescope of types which require only one universe to write, but for which nobody knows how to write a denotation function~\cite{shulman2014hott}.
We can write a function \mintinline{coq}{SST : ℕ → syntax}, and we can even write a function \mintinline{coq}{∀ n : ℕ, (SST n) is well-typed}, but nobody knows if it is possible to write a function \mintinline{coq}{ℕ → Type} interpreting this sequence of type syntax~\cite{kolomatskaia2022semisimplicial}.
Since semisimplicial types capture the essential idea of ``arbitrary many levels of type dependency,'' folklore conjectures that the problem of writing an interpreter for raw syntax with one universe, without assuming UIP (uniqueness of identity proofs) is \emph{equivalent} to the problem of internalizing semisimplicial types into a function \mintinline{coq}{ℕ → Type}.

While future results may allow us to bypass this obstacle entirely, autospecialization can again rescue us by determining how many levels of dependent judgmental equality are required, internalizing semisimplicial types on-the-fly up to this bound automatically---for example by unquoting the general syntax for semisimplicial types---and having the entire denotation function be parameterized on a ``partial semisimplicial gadget''.

\backmatter

\section*{Statements and Declarations}

\bmhead*{Supplementary information}

\newcommand{\archivedatswh}[2][]{archived at \\ \href{https://archive.softwareheritage.org/#2#1}{#2}}
Our code and data is available on GitHub:
\begin{itemize}
\item
  Our rewriting framework: \githublink{mit-plv}{rewriter} \archivedatswh[;origin=https://github.com/mit-plv/rewriter;visit=swh:1:snp:6d296391d575112bf64de359f5ee0efad7b89305]{swh:1:rev:f3f6bc1f48e4ff7475e0bba53c679e7a774114db}
\item
  Fiat Cryptography: \githublink{mit-plv}{fiat-crypto} \archivedatswh[;origin=https://github.com/mit-plv/fiat-crypto;visit=swh:1:snp:abfb67d6a0f7a0b268b74449951b08dbf8bf07f5]{swh:1:rev:8377bca1f8e2bdc7997480fd33b1492291f87c8c}
\item
  Experiments with Lean: \githublink{mit-plv}{fiat-crypto}[lean] \archivedatswh[;origin=https://github.com/mit-plv/fiat-crypto;visit=swh:1:snp:abfb67d6a0f7a0b268b74449951b08dbf8bf07f5;anchor=swh:1:rev:4a5bb46546a9a4dd276f62fc6be3fc7677e9f3c4;path=/fiat-crypto-lean/]{swh:1:dir:a5c9b1b2c700e061832d1cb87e9348bd2815c5e2}
\item
  Experiments mixing Fiat Cryptography with \tacsetoidrewrite{} and \tacrewritestrat{}: \githublink{coq-community}{coq-performance-tests} (in \texttt{src/fiat\_crypto\_via\_setoid\_rewrite\_standalone.v}) \archivedatswh[;origin=https://github.com/coq-community/coq-performance-tests;visit=swh:1:snp:cb6288953bb2cdc3026bc960d2f0227a5392d27b;anchor=swh:1:rev:4aaef74e5742fe3ad87c5d18d735e82b1284ecc6;path=/src/fiat_crypto_via_setoid_rewrite_standalone.v]{swh:1:cnt:9135903b92751770505441d8b011cdae01da3e7d}
\item
  Microbenchmarks performance evaluation: \\ \githublink{mit-plv}{rewriter}[ITP-2022-perf-data] \archivedatswh[;origin=https://github.com/mit-plv/rewriter;visit=swh:1:snp:6d296391d575112bf64de359f5ee0efad7b89305]{swh:1:rev:1787ab401a7e71afc9937010e2e155e4b1594ab5}
\item
  Fiat Cryptography performance evaluation: \\ \githublink{mit-plv}{fiat-crypto}[perf-testing-data-ITP-2022-rewriting] \archivedatswh[;origin=https://github.com/mit-plv/fiat-crypto;visit=swh:1:snp:6d8205c7c35b7b6fd767dbebd2089f4384ca4691]{swh:1:rev:72fe0dddee5e6dceeab0b8a2e6a745abf5287d3e}
\end{itemize}
More detail on the particular performance experiments we ran, as well as instructions for reading the version of the code supplement included with our ITP paper are available in the appendices of the arXiv version of our ITP submission~\cite{RewritingITP2022}.

\bmhead*{Funding}
\fundingtext
%\bmhead*{Acknowledgments}

%\todo{Acknowledgments are not compulsory. Where included they should be brief. Grant or contribution numbers may be acknowledged.
%
%Please refer to Journal-level guidance for any specific requirements.}

%This work was supported in part by a Google Research Award, National Science Foundation grants CCF-1253229, CCF-1512611, and CCF-1521584, and %the National Science Foundation Graduate Research Fellowship under Grant Nos.\ 1122374 and 1745302.
%Any opinion, findings, and conclusions or recommendations expressed in this material are those of the authors and do not necessarily reflect the views of the National Science Foundation.%

%Some journals require declarations to be submitted in a standardised format. Please check the Instructions for Authors of the journal to which you are submitting to see if you need to complete this section. If yes, your manuscript must contain the following sections under the heading `Declarations':
%If any of the sections are not relevant to your manuscript, please include the heading and write `Not applicable' for that section.

\bmhead*{Competing Interests}
\paragraph*{Financial interests}
None

\paragraph*{Non-financial interests}
Jason Gross is a member of the Coq development team.

\bmhead*{Ethics approval}
Not applicable
\bmhead*{Consent to participate}
Not applicable
\bmhead*{Consent for publication}
Not applicable
%\bmhead*{Availability of data and materials}
%Not applicable
%\bmhead*{Code availability}
%Not applicable
\bmhead*{Authors' contributions}

Jason Gross wrote the bulk of the code of the rewriting framework---building on a proof-of-concept prototype Andres Erbsen wrote for Fiat Cryptography---performed the performance evaluations, and did some work on Fiat Cryptography.
Jason Gross, Andres Erbsen, Jade Philipoom, and Adam Chlipala contributed to the design of the rewriting framework.
Andres Erbsen, Jason Gross, and Jade Philipoom contributed to the integration of the rewriting framework with Fiat Cryptography and did non-performance evaluation and testing.

Jason Gross prepared all performance plots, wrote the technical explanations of the rewriting framework.
Jason Gross, Andres Erbsen, and Adam Chlipala wrote the technical content already present in the ITP submission.
All authors reviewed the original text of the ITP submission.
Jason Gross and Rajashree Agrawal, with input from and in conversation with Andres Erbsen, developed the new context of the rewriting framework as a prototype for a performant proof engine building block.
Jason Gross and Rajashree Agrawal wrote most of the new text of the article, with the section on the theoretical asymptotic analysis of incremental rewriting based on a draft by Andres Erbsen.

%\noindent

%%===================================================%%
%% For presentation purpose, we have included        %%
%% \bigskip command. please ignore this.             %%
%%===================================================%%
%\bigskip
%\begin{flushleft}%
%Editorial Policies for:
%
%\bigskip\noindent
%Springer journals and proceedings: \url{https://www.springer.com/gp/editorial-policies}
%
%\bigskip\noindent
%Nature Portfolio journals: \url{https://www.nature.com/nature-research/editorial-policies}
%
%\bigskip\noindent
%\textit{Scientific Reports}: \url{https://www.nature.com/srep/journal-policies/editorial-policies}
%
%\bigskip\noindent
%BMC journals: \url{https://www.biomedcentral.com/getpublished/editorial-policies}
%\end{flushleft}

%\url{https://www.cis.upenn.edu/~bcpierce/papers/binders.pdf}
%\href{https://www.cis.upenn.edu/~bcpierce/papers/binders.pdf}{https://www.cis.upenn.edu/~bcpierce/papers/binders.pdf}
%%===========================================================================================%%
%% If you are submitting to one of the Nature Portfolio journals, using the eJP submission   %%
%% system, please include the references within the manuscript file itself. You may do this  %%
%% by copying the reference list from your .bbl file, paste it into the main manuscript .tex %%
%% file, and delete the associated \verb+\bibliography+ commands.                            %%
%%===========================================================================================%%
\nocite{Schropp2013}
%\todo{report broken bib style around tildes (bcpierce), missing url on article, missing doi on inproceedings, inconsistent captialization, ``Warning--... isn't a brace-balanced string for entry RewritingITP2022 while executing--line 2823 of file sn-mathphys.bst'' misc to journal}
% hack to fix ~ in urls
\newcommand{\burl}[1]{{\edef\texttildelow{\detokenize{~}}\url{#1}}}
\bibliography{rewriting}% common bib file
%% if required, the content of .bbl file can be included here once bbl is generated
%%\input sn-article.bbl

%% Default %%
%%\input sn-sample-bib.tex%

\end{document}